\documentclass[3p,onecolumn]{elsarticle}

\usepackage{amsmath}
\usepackage{bm}

\usepackage{hyperref,booktabs}
\usepackage{xcolor}
\usepackage{multirow}
\usepackage{tikz}
\usetikzlibrary{calc}
\usetikzlibrary{shapes.geometric}
\usetikzlibrary {arrows.meta} 
 
\biboptions{sort&compress}

\newcommand{\refgeneral}[1]{#1}
\newcommand{\refthree}[1]{#1}

\journal{computer Physics Communications}

\begin{document}

\begin{frontmatter}

\title{
Multi-GPU Acceleration of PALABOS Fluid Solver using C++ Standard Parallelism
}

\author[unige]{Jonas Latt\corref{correspondingauthor}}
\cortext[correspondingauthor]{Corresponding author}
\ead{jonas.latt@unige.ch}
\author[bnbu,unige]{Christophe Coreixas}
\address[unige]{Department of Computer Science, University of Geneva, 1227 Carouge, Switzerland.}
\address[bnbu]{Institute for Advanced Study, Beijing Normal - Hong Kong Baptist University, Zhuhai 519088, China.}

\begin{abstract}
This article presents the principles, software architecture, and performance analysis of the GPU port of the lattice Boltzmann software library Palabos \href{https://doi.org/10.1016/j.camwa.2020.03.022}{[J. Latt et al., ``Palabos: Parallel lattice Boltzmann solver", Comput. Math. Appl. 81, 334–350, (2021)]}. A hybrid CPU-GPU execution model is adopted, in which numerical components are selectively assigned to either the CPU or the GPU, depending on considerations of performance or convenience. This design enables a progressive porting strategy, allowing most features of the original CPU-based codebase to be gradually and seamlessly adapted to GPU execution.
The new architecture builds upon two complementary paradigms: a classical object-oriented structure for CPU execution, and a data-oriented counterpart for GPUs, which reproduces the modularity of the original code while eliminating object-oriented overhead detrimental to GPU performance. Central to this approach is the use of modern C++, including standard parallel algorithms and template metaprogramming techniques, which permit the generation of hardware-agnostic computational kernels. This facilitates the development of user-defined, GPU-accelerated components such as collision operators or boundary conditions, while preserving compatibility with the existing codebase and avoiding the need for external libraries or non-standard language extensions.
The correctness and performance of the GPU-enabled Palabos are demonstrated through a series of three-dimensional multiphysics benchmarks, including the laminar–turbulent transition in a Taylor–Green vortex, lid-driven cavity flow, and pore-scale flow in Berea sandstone. Despite the high-level abstraction of the implementation,  
the single-GPU performance is similar to CUDA-native solvers, and multi-GPU tests exhibit good weak and strong scaling across all test cases.
Beyond the specific context of Palabos, the porting methodology illustrated here provides a generalizable framework for adapting large, complex C++ simulation codes to GPU architectures, while maintaining extensibility, maintainability, and high computational performance.
\\


\noindent \textbf{PROGRAM SUMMARY}

\begin{small}
\noindent
{\em Program Title:} Palabos: Parallel Lattice Boltzmann Solver \\
{\em CPC Library link to program files:} (to be added by Technical Editor) \\
{\em Developer's repository link:} \refgeneral{\url{https://gitlab.com/unigespc/palabos}} \\
{\em Licensing provisions:} AGPLv3 \\
{\em Programming language:} C++17, using C++ parallel algorithms \\
{\em Supplementary material:} A compressed file containing a snapshot of the described code version \\
{\em Nature of problem:} \\
The lattice Boltzmann method (LBM) is a widely used approach in computational fluid dynamics (CFD) for simulating flows in complex geometries, turbulence, multiphase systems, and thermal transport. The Palabos library was designed to provide the scientific community with a flexible and efficient parallel LBM platform, implementing a broad range of physical and numerical models. Until now, Palabos has been CPU-based, relying on distributed-memory parallelism through MPI. However, with modern simulation platforms and supercomputers increasingly dominated by GPU accelerators, a CPU-only implementation no longer provides competitive performance. This work presents a complete redesign of the Palabos architecture to run efficiently on multi-GPU systems. The new code supports hybrid execution: legacy CPU components remain functional, new GPU-optimized modules can be developed within a compatible programming model, and both can interact seamlessly. This version supersedes the original CPU-only implementation and will serve as the basis for all future development of Palabos. \\

{\em Solution method:} \\
The GPU version of Palabos introduces a new data container, the \texttt{AcceleratedLattice}, designed specifically for accelerator execution. It preserves the user interface of the CPU-based \texttt{MultiBlockLattice}, but replaces its array-of-structures layout and virtual-function polymorphism with a structure-of-arrays layout, thread-safe collision–streaming schemes, and integer tags for model selection. These changes enable efficient memory access, fine-grained parallelism, and compatibility with GPU execution models, while maintaining backward compatibility and supporting hybrid CPU/GPU workflows.  

Parallelism is expressed through C++17 standard parallel algorithms, compiled with the NVIDIA HPC SDK. Fundamental parallel algorithms such as \texttt{for\_each}, \texttt{transform\_reduce}, and \texttt{exclusive\_scan} are used to implement the essential building blocks of the lattice Boltzmann method, including linear processing, reductions, and data packing for MPI communication. This data-oriented design yields a portable and maintainable GPU implementation, preserves modularity, and ensures that existing Palabos applications can be ported to GPU with minimal modifications. \\

{\em Additional comments including restrictions and unusual features:} \\
The GPU version currently relies on the platform-independent framework of C++ parallel algorithms but has, as of now, only been tested with NVIDIA GPUs (compiled with nvc++) supports NVIDIA architectures with CUDA-compatible compilers (tested with nvc++). The current version of the code does not compile with the 25.X series of the nvc++ compiler. \\

\end{small}

\end{abstract}


\end{frontmatter}

\section{Introduction\label{sec:intro}}
Porting computationally intensive scientific software to accelerated hardware can lead to substantial performance improvements. Typical devices used to accelerate scientific applications include field-programmable gate arrays (FPGAs), domain-specific accelerators such as tensor-processing units (TPUs), and Graphics Processing Units (GPUs). The latter have proved particularly successful due to their affordability and relatively broad range of use. Compared to CPUs, GPUs typically offer a substantially higher peak rate of floating-point computations, and a higher peak memory bandwidth, at an equivalent financial cost or energy consumption. Conversely, GPUs can be more challenging to use due to a more restrictive programming model. 

General-purpose GPU (GPGPU) programming languages provide a low-level programming interface that is closely tied to the GPU architecture.
For instance, CUDA~\cite{NICKOLLS_QUEUE_2008} is designed for developing applications that run efficiently on NVIDIA GPUs only, while HIP~\cite{HIP_WEBSITE} is more commonly used for AMD GPUs.
Alternatively, higher-level approaches have emerged, offering hardware-agnostic solutions that allow, e.g., the same code to run on CPUs and GPUs provided by different vendors. Examples include directive-based coding strategies (OpenACC~\cite{WIENKE_EUROPAR_2012} and OpenMP~\cite{CHANDRA_Book_2001}) and multi-architecture libraries/frameworks (OpenCL~\cite{MUNSHI_IEEE_2009}, SYCL~\cite{ALPAY_PIWOCL_2020}, Thrust~\cite{HWU_Chapter_2012}, Kokkos~\cite{CARTEREDWRADS_JPDC_74_2014}). These approaches enable parts of the code to be offloaded to the GPU with minimal modifications to the original CPU code.
A more recent strategy consists in using parallel features available in ISO languages such as C++ and Fortran. This strategy simplifies development and enhances portability by providing a uniform way to write code for both standard CPUs and accelerated hardware. In that context, switching to different target hardware only requires a change of compiler or compilation flag~\cite{LARKIN_GTC_2022,LARKIN_GTC_2024}.

The available C++ parallel algorithms efficiently implement elementary operations, such as search or sorting tasks. Furthermore, generic data-processing algorithms like \texttt{for\_each} and \texttt{transform\_reduce} are available, which are customized through user-defined functions and offer a functionality equivalent to a parallelized \texttt{for} loop with reductions. These algorithms allow solving a broad range of problems, including stencil operations on a structured grid. We demonstrated this point in \cite{LATT_PLOSONE_16_2021}, which uses C++ parallel algorithms to execute a Computational Fluid Dynamics (CFD) simulation on GPU. The provided open-source demonstration code, based on the lattice Boltzmann method (LBM) and using a homogeneous grid, runs at high performance on GPU. Since the original code implementation, the performance gap of 20\% has been narrowed to only a few percent compared to a reference code written in the CUDA programming language. Interestingly, this code demonstrated the feasibility of implementing state-of-the-art scientific software with limited GPU-specific knowledge, and with a portable approach that is exempt from any GPU vendor lock-in and even applies to multiple types of accelerators. Since then, ISO-standard languages have been increasingly adopted to accelerate a wide range of simulation tools on GPUs from various vendors (NVIDIA, AMD, and Intel). These efforts include the acceleration of high-order finite-difference methods~\cite{COREIXAS_HIFILED_LBFDGPU_2022,CAPLAN_AJSS_278_2025,CAPLAN_SC_2025}, semi-Lagrangian schemes~\cite{ASAHI_IEEE_2022}, discrete element methods~\cite{MAGGIOAPRILE_Master_2023,RAMBOSSON_Master_2024,MAGGIOAPRILE_UnderPreparation_2025}), mini-applications~\cite{LIN_IEEE_2022,HASEEB_SC_2023,LIN_ARXIV_02680_2024}, and even LB codes running on non-uniform grids~\cite{COREIXAS_ARXIV_04465_2025}.

The present article builds on this procedure and proposes a methodology to port an existing CPU-based C++ software library to GPU using parallel algorithms. 
Particular attention is devoted to the MPI communication layer of the original CPU base, which is reused to allow multi-GPU execution. The resulting code is tested using examples of the LBM software library Palabos~\cite{LATT_CMA_81_2021}, many of which were adapted for an execution on a multi-GPU environment prior to drafting this article~\cite{LATT_NVIDIA_BLOG_1_2022,LATT_NVIDIA_BLOG_2_2022}. Despite the wide scope of the Palabos library~\cite{LAGRAVA_JCP_231_2012,BROGI_JASA_142_2017,LYU_PoF_35_2023,PARMIGIANI_IJMPC_24_2013,MORRISON_CF_172_2018,MERCIER_AOR_97_2020,GRONDEAU_EN_15_2022,LOLIES_AMS_2019,SEIL_ICDEM_LBDEM_2017,PARMIGIANI_NATURE_532_2016,LEMUS_FRONTIERS_9_2021,LECLAIRE_IJMPC_28_2017,ANVARI_SR_11_2021,SHODIEV_ESM_38_2021,KOTSALOS_JCP_398_2019,PETKANCHIN_SR_13_2023,CONRADIN_JCS_53_2021}, the development effort of the GPU port was accomplished within a few months as a side project~\cite{LATT_DSFD_GPU_2021} and without the assignation of a dedicated developer. This illustrates the light burden of the proposed method, and the adequacy of the method for developers with limited GPU programming knowledge. The characteristics of the GPU port achieved with our method are summarized as follows:
\begin{itemize}
    \item The existing API of the CPU is maintained to allow previously developed applications to run on both CPU and GPU backends.
    \item Hybrid CPU/GPU execution is allowed to guarantee access to software components that have not yet been ported or cannot be ported to GPU. Multi-node MPI parallelism is implemented for both CPU and GPU components in a compatible manner to allow fast data exchange.
    \item Code duplication is minimal, as a common code base is maintained for both CPU and GPU executions.
\end{itemize}

The rest of the paper is organized as follows. First, Section~\ref{sec:description} describes the GPU porting strategy and its application to Palabos fluid solver. Next, Section~\ref{sec:validation} validates the GPU version of the fluid solver through several benchmarks relevant to its users. Section~\ref{sec:perfo} evaluates its performance on a single A100-SXM4 GPU (40GB) and a full DGX-A100-SXM4 (4$\times$40GB). Finally, Section~\ref{sec:conclusion} presents the conclusions regarding the multi-GPU porting of Palabos fluid solver.

All numerical tests in this article can be reproduced using the pre-release version of the GPU-ported Palabos version on the Palabos Web page \url{https://palabos.unige.ch}. The test cases are found in the \texttt{examples/gpuExamples} directory.

\section{Description of the approach\label{sec:description}}

\subsection{Overview}
In the original version of Palabos, the so-called \texttt{MultiBlock} serves as a fundamental container for all parallelized, large data sets. One of its instances, the \texttt{MultiBlockLattice}, holds the LBM populations, which are regrouped using a blockwise domain decomposition and assigned to multiple CPU cores and/or multi-processor nodes using MPI primitives. The \texttt{MultiBlockLattice} offers a flexible user interface, because (1) the number of populations and lattice connectivity can be adjusted using template metaprogramming, via \emph{lattice descriptors}, (2) the LBM collision model can be adjusted locally using an object-oriented runtime mechanism, through \emph{dynamics objects}, and (3) non-local algorithm ingredients, like the execution of finite-difference stencils, are expressed using the formalism of Palabos \emph{data processors} (additional information is provided in the main Palabos reference~\cite{LATT_CMA_81_2021}). Because of the data organization and execution-model issues described in Section~\ref{sec:gpu-implementation}, accessing the data and associated functionality of the \texttt{MultiBlockLattice} directly from within a GPU leads to insurmountable performance issues and even, for some aspects, to technical obstacles that could not be solved with current compiler technology and GPU architecture. Instead, Palabos was enhanced with a new data container, the \texttt{AcceleratedLattice}.

To maintain backward compatibility, the \texttt{AcceleratedLattice} provides the same user interface as the \texttt{MultiBlockLattice}, and in particular, its lattice connectivity can be adjusted using the same lattice descriptors. However, for reasons explained in Section~\ref{sec:gpu-implementation}, it relies on its own implementation of local collision models (formerly \emph{dynamics objects}) and non-local stencils (\emph{data processors}). 
A large number of such \emph{dynamics objects} and \emph{data processors} have been ported to their GPU counterparts to accompany this GPU release of Palabos, and a formalism has been introduced to support further migration of existing Palabos extensions in a largely automated manner. Furthermore, bridging mechanisms were provided to exchange data between the \texttt{MultiBlockLattice} and \texttt{AcceleratedLattice} in run time, thus allowing for a hybrid CPU/GPU execution model.

\begin{figure}[hbtp]
    \centering
    \includegraphics[width=\textwidth]{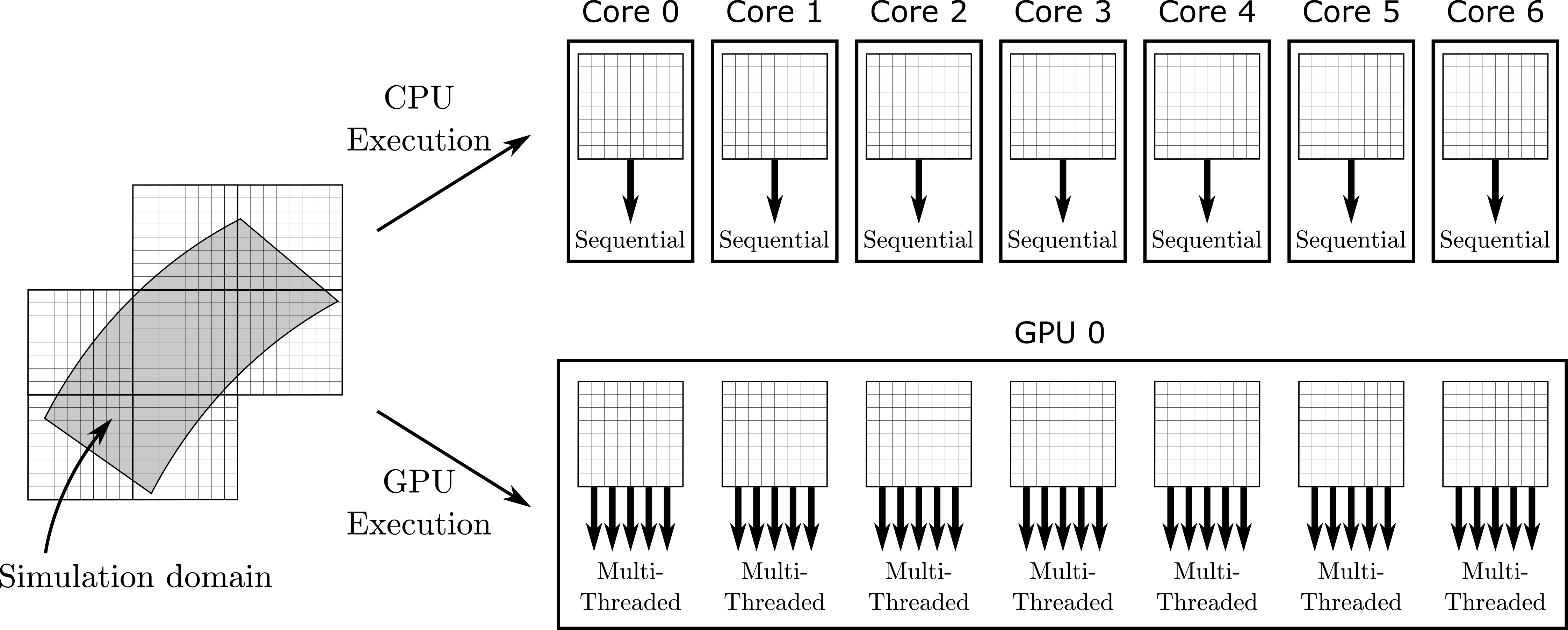}
    \caption{The grid covering the simulation domain is partitioned into regular blocks. On the CPU, each block is processed sequentially by its assigned CPU core. On the GPU, each block is processed in parallel through the shared-memory formalism of parallel algorithms.}
    \label{fig:domain_repartition}
\end{figure}
Figure~\ref{fig:domain_repartition} illustrates the different parallelization strategies applied in the two Palabos data containers, both of which rely on domain decomposition into regular blocks (left part). On the CPU (upper right part), each CPU core processes one or more blocks sequentially, allowing non-local data accesses to be efficiently resolved by a matrix indexing scheme. The granularity of the blocks is determined by the efficiency trade-off between allocating unused cells in large blocks and the need for excessive data transfer in small blocks (see~\cite{LATT_CMA_81_2021} for more details). This approach is however inefficient on GPUs, which typically operate with a larger number of execution threads and thus rely on a finer-grained form of parallelism. Therefore, the operations executed within a block are further parallelized within the GPU (lower right part) using the C++ formalism of parallel algorithms described in Section~\ref{sec:programming-model}. Optimal performance on the GPU is typically achieved by relying on decomposition into larger blocks than on the CPU, so that only a few blocks, or even a single one, are assigned to each GPU.
However, despite this different execution mode, both data containers rely on the same MPI-based interblock communication layer. It should be mentioned that the formalism of C++ parallel algorithms can be applied to different types of homogeneous and heterogeneous parallel platforms, although this article focuses on GPUs. In particular, the \texttt{AcceleratedLattice} can be used to implement hybrid MPI -- shared memory parallelization by applying parallel algorithms to a multi-core CPU. In that case, both a dedicated compiler (e.g. nvc++) or a general-purpose compiler with Intel's Threading Building Blocks library~\cite{KIM_IEEE_2011} can be used to run an \texttt{AcceleratedLattice} on multi-core CPUs.

\refthree{We finally comment on the learning curve associated with the Palabos GPU port. For applications that rely exclusively on functionality already available in the GPU-enabled code base, most notably standard lattice Boltzmann models and commonly used data processor, the transition is essentially immediate, as it suffices to replace the \texttt{MultiBlockLattice} by an \texttt{AcceleratedLattice}. For users who have developed custom extensions, such as application-specific dynamics objects or data processors, an explicit port to the new GPU-oriented formalism is required, following the patterns provided by the existing examples. While this entails a non-negligible initial effort, the learning curve is significantly softened by the hybrid design of Palabos, which allows functionalities to be ported incrementally and validated step by step, rather than requiring a full rewrite of the application.}

In summary, Palabos-based end-user applications are easily ported to GPU by replacing all reference to a \texttt{MultiBlockLattice} to an \texttt{AcceleratedLattice}, and by using the ported version of all data processors that are manually invoked, such as finite-difference stencil operators. Palabos also supports a hybrid execution model in which both a \texttt{MultiBlockLattice} and an \texttt{AcceleratedLattice} data container are instantiated to mirror the simulation results on CPU and GPU, respectively. The data can be copied between either container on a per-need basis to proceed with the computations on the hardware platform on which better suited.

\subsection{Programming Model}\label{sec:programming-model}
The GPU port of Palabos is realized using C++ standard language parallelism. Available since the C++17 language standard, this hardware agnostic formalism allows compilers to parallelize algorithms of the C++ Standard Template Library (STL) using a shared-memory paradigm chosen by the compiler. Current compiler implementations include GPU and multi-core CPU parallelism. For LBM applications, the formalism showed performance close to that expected from other high-level formalisms, with only a small trade-off in efficiency compared to lower-level approaches such as CUDA~\cite{LATT_PLOSONE_16_2021,COREIXAS_ARXIV_04465_2025}. The choice of this approach is driven by portability, as it relies solely on intrinsic, ISO-certified language features, and convenience. Thanks to the language integration of this parallel formalism, adapting a code for GPU execution can often be achieved with minimal or no syntactic changes. Nonetheless, a solid understanding of the GPU execution model remains crucial, as discussed in Section~\ref{sec:gpu-implementation}.

The fundamental workhorse for a standard-language parallelization of an LBM-like code are the STL algorithms \texttt{for\_each} and \texttt{transform\_reduce}, which essentially express a functional-style \emph{for} loop, with (\texttt{transform\_reduce}) or without (\texttt{for\_each}) reductions. The behavior of the \texttt{for\_each} algorithm can be adjusted by a user-defined operation, equivalent to the body of a \emph{for} loop. This operation is distributed across multiple execution threads when the parallel version of the algorithm is chosen, similar to the effect of parallelizing a \emph{for} loop with the OpenMP formalism. However, the strength of this approach lies in the availability of numerous parallelized algorithms, including algorithms that rely on data dependency between threads. More precisely, the GPU port of Palabos relies on the following algorithms.

\paragraph{Linear Processing}
The \texttt{for\_each} algorithm is used to process data linearly and in-place. Cases where the output container is distinct from the input, the \texttt{transform} algorithm is commonly used instead, as a matter of syntactic preference.

\paragraph{Reductions}
The \texttt{reduce} and \texttt{transform\_reduce} algorithms are used for all types of reduction operations. Note that due to the severe performance penalty of repeated memory traversals, a \texttt{transform\_reduce} call is commonly used in Palabos to express both a linear data processing and a reduction operation in a single call.

\paragraph{Prefix Sums}
The \texttt{exclusive\_scan} and \texttt{transform\_exclusive\_scan} algorithms are applied to all types of re-indexing operations, i.e., whenever data is mapped into a larger or smaller container at different index locations. An example includes data packing and unpacking operations for MPI communication, which must be carried out within GPU memory to guarantee good parallel performance on multi-GPU systems that support direct GPU-to-GPU communication.

\subsection{Code design for GPU execution}\label{sec:gpu-implementation}
To port the original Palabos to GPU, it can seem sufficient to apply STL algorithms to parallelize all for loops appearing in its code base. However, this approach falls short due to limitations in data layout, memory access patterns, and execution models, which can result in suboptimal performance or even prevent the code from running on GPU altogether. To properly address the needs of a GPU execution, we introduce a new data container, the \texttt{AcceleratedLattice}, designed specifically for GPU execution, alongside the existing \texttt{MultiBlockLattice}. This new structure promotes uniformity in both data access and execution flow. While the original Palabos framework follows an object-oriented paradigm, we use the terminology of \emph{data-oriented design} to refer to the GPU-optimized version. This shift places primary emphasis on how data is organized and accessed, with higher-level abstractions introduced only afterward to ensure maintainability and adherence to sound software engineering principles. The fundamental ideas of this design are described in this section.

\subsubsection{Structure-of-array memory layout}
One of the key differences between the \texttt{AcceleratedLattice} and \texttt{MultiBlockLattice} containers lies in the choice of data structure. The \texttt{MultiBlockLattice} chooses an array-of-structure (AoS) layout, which is well suited to the object-oriented approach of the original Palabos code, because it places all data belonging to the same grid node at successive memory addresses. Nonetheless, it is well known~\cite{TOLKE_CVS_13_2010} that a structure-of-array (soa) layout is better suited for GPU-based LBM implementations. Our previous work~\cite{LATT_PLOSONE_16_2021} notably shows that within the formalism of C++, a four-fold (or even higher) speedup can be achieved by switching from AoS to soa.

\begin{figure}[b]
    \centering
    \includegraphics[width=.9\textwidth]{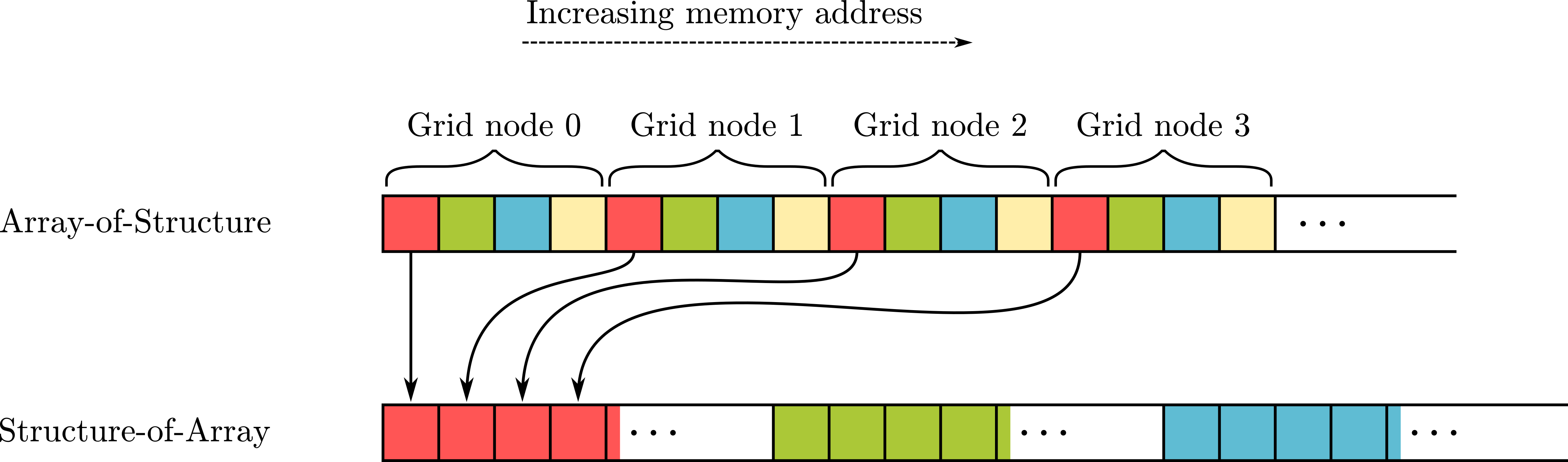}
    \caption{Array-of-Structure versus Structure-of-Array data layout in an example with four data elements per grid node.}
    \label{fig:aos_vs_soa}
\end{figure}
Figure~\ref{fig:aos_vs_soa} illustrates the two implementation strategies in a fictional example with four data elements per grid node, using a different color for each type of data element (or population index). In the LBM, this number is actually larger, with typically 19 or 27 populations. To understand the performance difference between the two approaches, it is necessary to focus on the streaming step, which is less naturally inclined to achieve optimal performance due to the complex memory access pattern. As shown in Figure~\ref{fig:aos_vs_soa}, an SoA implementation strategy groups all variables that communicate during the streaming step (e.g., all red cells or all $f_1$ populations) into a contiguous block. Thanks to this, the regularity of the memory access scheme is greatly enhanced, thus explaining the impressive GPU speedups obtained with the SoA data layout for LBMs.
On CPUs, however, the situation is less clear-cut, and the preferred data layout depends on the type of CPU and the number of cores~\cite{LATT_PLOSONE_16_2021}.

It is also worth noting that the \texttt{MultiBlockLattice} executes the collision-streaming cycles in a thread-unsafe manner within a single block using a fused in-place swap algorithm, which is unproblematic in the single-threaded execution per-block operations of the original Palabos library. For the multi-threaded execution model of the \texttt{AcceleratedLattice}, it is necessary to switch to a thread-safe approach, several of which (two-population, AA-pattern, swap with separate collision and streaming, esoteric twists) are presented in Refs.~\cite{LATT_PLOSONE_16_2021,HOLZER_PhD_2025}. The current GPU port of Palabos uses the two-population scheme, which doubles the memory requirements for the populations but guarantees performances compatible with other approaches. Future work will consider switching to a non-fused swap approach, such as the AA pattern, to halve the memory footprint without significantly impacting performance, albeit at the cost of a more complex implementation.

\subsubsection{Collision-model polymorphism without virtual function calls}
The \texttt{MultiBlockLattice} in Palabos adopts a natural object-oriented model in which every grid node is represented by an object that regroups all data attributed to the node, including the populations. The collision step is executed polymorphically through virtual function-call resolution, allowing the collision to be adjusted locally during run time. Furthermore, Palabos users can design user-specific collision terms by designing their own class of type \texttt{Dynamics}, allowing to freely adjust the behavior of their simulation on a per-cell basis. This choice caters to the typical needs of the LBM community by allowing variable physical and numerical properties in space and time, and defines the versatility of the Palabos library. However, it presents several challenges for the GPU port. The first issue arises from the incompatibility of data layouts: the AoS layout naturally adopted in an object-oriented approach conflicts with the SoA layout required for efficient GPU implementation. Secondly, virtual functional-call resolution did not appear as an acceptable mechanism for the GPU port, as it was not supported by the compiler of the NVIDIA HPC SDK used for the tests and benchmarks of this work. Even if this technique was supported in future compiler versions, it would fundamentally conflict with performance considerations on the rather simple execution model of GPUs.

Switching from an AoS to an SoA data layout for the \texttt{AcceleratedLattice} presents a challenge for code reuse. This is particularly true for the significant portion of the Palabos code handling various collision models and other cell processing steps.
This problem was overcome using a classical push-pull implementation of the collision and streaming steps. This principle implies that the data of a node that is scattered in GPU memory following the SoA layout of the \texttt{AcceleratedLattice} is pulled into local, adjacent variables whenever a node is processed in the collision-streaming cycle, after which original Palabos code portions are applied to the local node. In such a way, unnecessary code duplication between the original and GPU-ported parts of Palabos is avoided.

As a replacement for the virtual function-call approach in \texttt{MultiBlockLattice}, the \texttt{AcceleratedLattice} assigns an integer tag to every node, which is used to invoke the appropriate collision model using a \texttt{switch}-like language construct.  The implemented formalism provides a one-to-one correspondence between the new tag and the original \texttt{Dynamics} object, to allow the \texttt{MultiBlockLattice} and the \texttt{AcceleratedLattice} to mirror each other for pre- and post-processing tasks, or for a CPU/GPU-hybrid program execution. In this context, a run-time type identification (RTTI) mechanism is required, both to assign a unique identifier to each collision model and to allow automatic instantiation of the respective object-oriented and data-oriented structures. Given the limited RTTI support of the C++ language, our method relies on additional library-level RTTI functionality. In Palabos, the mentioned unique integer tags are complemented with string identifiers, which were originally implemented for network transfer and checkpointing of serialized data (see Ref.~\cite{LATT_CMA_81_2021}). 
Depending on the context, different identification methods may be preferred. Tag-based identification is more efficient and suited for tasks like collision-streaming kernel execution, while string-based identification is more user-friendly and used in cases such as custom kernel generation (see Section~\ref{sec:kernel-size-reduction}).

\begin{figure}
    \centering
    \begin{tikzpicture}[scale=0.5]
        \def\rectwidth{3cm}
        \def\rectheight{0.5cm}
        \def\spacing{3cm}
        \def\linewidth{0.25mm}
        \def\x0{0.5}
        \node[draw, rounded corners, minimum width=\rectwidth, minimum height=\rectheight, line width=\linewidth, align=center, font=\small] (rect1) at (\x0,0) {Left boundary};
        \node[draw, rounded corners, minimum width=2.5cm, minimum height=\rectheight, line width=\linewidth, align=center, font=\small] (rect2) at ($(rect1.south) + (0, -0.5*\spacing)$) {Smagorinsky};
        \node[draw, rounded corners, minimum width=1.5*\rectheight, minimum height=\rectheight, line width=\linewidth, align=center, font=\small] (rect3) at ($(rect2.south) + (0, -0.5*\spacing)$) {RR};
        \node[draw, rounded corners, minimum width=\rectheight, minimum height=\rectheight, line width=\linewidth, align=center, font=\small] (rect4) at ($(rect3.south) + (0, -0.5*\spacing)$) {\:\circle*{2}};
        \node[draw, ellipse, minimum width=1.5*\rectheight, minimum height=\rectheight, line width=\linewidth, align=center, font=\small] (ellips) at ($(rect1.south) + (4*\spacing, -0.5*\spacing)$) {Integer tag};
        \node[draw, rounded corners, minimum width=1.5*\rectheight, minimum height=\rectheight, line width=\linewidth, align=center, font=\small] (rect5) at ($(rect2.south) + (6*\spacing, -0.5*\spacing)$) {Merged code};
        \node[draw, rounded corners, minimum width=\rectheight, minimum height=\rectheight, line width=\linewidth, align=center, font=\small] (rect6) at ($(rect5.south) + (0, -0.5*\spacing)$) {\:\circle*{2}};
        \node[above, align=center, font=\bfseries\normalsize](obj) at (\x0,0.35*\spacing) {Object-oriented}; 
        \node[left, font=\small, align=right] at ($(rect1) + (-1.1*\spacing, 0)$) {Boundary-completion class};
        \node[left, font=\small, align=right] at ($(rect2) + (-1.1*\spacing, 0)$) {LES-model class};
        \node[left, font=\small, align=right] at ($(rect3) + (-1.1*\spacing, 0)$) {Collision-model class};
        \node[left, font=\small, align=right] at ($(rect4) + (-1.1*\spacing, 0)$) {Grid node};
        \node[align=center, font=\bfseries\normalsize] at ($(obj) + (6.0*\spacing, 0)$) {SOA / Node Tagging};
        \draw[-stealth, thick] (rect1.south) -- (rect2.north);
        \draw[-stealth, thick] (rect2.south) -- (rect3.north);
        \draw[-stealth, thick] (rect3.south) -- (rect4.north);
        \draw[-stealth, thick] (rect1.east) to[out=340,in=180] node[above, font=\footnotesize\color{red!60!black}, rotate=-10] {BC\_Left} ($(ellips.west) + (-0.2,+1.0)$);
        \draw[-stealth, thick] (rect2.east) -- node[above, font=\footnotesize\color{red!60!black}] {LES\_Smagorinsky} ($(ellips.west) + (-0.2,0.0)$);
        \draw[-stealth, thick] (rect3.east) to[out=20,in=180] node[above, font=\footnotesize\color{red!60!black}, rotate=5] {COLL\_RR} ($(ellips.west) + (-0.2,-1.0)$);
        \draw[-stealth, thick] (rect5.south) -- (rect6.north);
        \draw[-stealth, thick] (ellips.south) to[out=300,in=180] ($(rect5.west) + (-0.1,0.0)$);
    \end{tikzpicture}
    \caption{A succession of polymorphic objects in the object-oriented structure translates to a unique integer identifier for the data-oriented data structure, exploiting Palabos's built-in unique string identification of numerical model classes.}
    \label{fig:from_oo_to_do}
\end{figure}
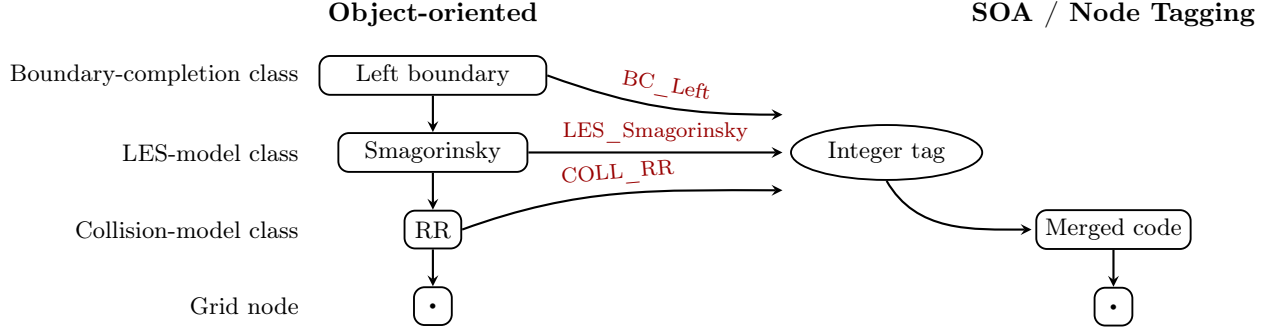
In Palabos, a specific difficulty stems from the handling of so-called \emph{chained collision models} that are frequently used in a \texttt{MultiBlockLattice} and arise when a chain of events is composed through a sequence of \texttt{Dynamics} objects that reference each other in a linked-list fashion. A grid node may for example implement a given basic collision model, such as the recursive-regularized (RR) collision~\cite{MALASPINAS_ARXIV_2015,COREIXAS_PRE_100_2019}, but also rely on a turbulence model like the Smagorinsky subgrid-scale model~\cite{HOU_ARXIV_1994,MALASPINAS_JFM_700_2012}, and further depend on an added completion algorithm for unknown variables on a boundary node. In the \textsf{AcceleratedLattice}, this chain must be identified through a unique tag, which is automatically assigned by the GPU framework of Palabos. The integer tag is deduced from the index of an ordered list of string identifiers, which themselves are  concatenated versions of the string identifiers of individual model components, as defined by the object-oriented Palabos version. The underlying mechanism is illustrated on Figure~\ref{fig:from_oo_to_do}.

A final aspect tackled by the Palabos framework responsible for the correspondence between the \textsf{MultiBlockLattice} and the \textsf{AcceleratedLattices}'s polymorphic behavior is the storage of collision-specific model parameters. For example, a node implementing a Smagorinsky subgrid-scale LES closure needs access to the Samgorinsky constant, which, in the original version of Palabos, is stored within the corresponding \texttt{Dynamics} object. Consistently with the SoA vision of the new GPU framework, the \textsf{AcceleratedLattice} allocates a separate array for the model parameters of all nodes. Additionally to the integer tag, the nodes now store an index into this array to access additional dynamic content on a per-need basis.
\begin{figure}
    \centering
    \begin{tikzpicture}
        \def\rectwidth{5cm}
        \def\rectheight{3cm}
        \def\spacing{3cm}
        \def\linewidth{0.5mm}
        \node[draw, rounded corners, minimum width=1.1*\rectwidth, minimum height=0.5*\rectheight, line width=\linewidth, align=left, font=\itshape\small] (rect1) at (0,0) {- Tagging of polymorphic objects \\ - Extraction of model parameters \\ - Multi-GPU communication policy};
        \node[draw, rounded corners, minimum width=0.9*\rectwidth, minimum height=1.1*\rectheight, line width=\linewidth, align=center, font=\small] (rect2) at ($(rect1.west) + (-\spacing, -1.2*\spacing)$) {};
        \node[draw, minimum width=0.5*\rectwidth, minimum height=0.2*\rectheight, line width=\linewidth, align=center, font=\small] (rect21) at ($(rect2) + (-0.2*\spacing, 0.35*\spacing)$) {MPI or OpenMP};
        \node[draw, minimum width=0.5*\rectwidth, minimum height=0.2*\rectheight, line width=\linewidth, align=center, font=\small] (rect22) at ($(rect2) + (-0.2*\spacing, 0.1*\spacing)$) {MPI or OpenMP};
        \node[draw, minimum width=0.5*\rectwidth, minimum height=0.2*\rectheight, line width=\linewidth, align=center, font=\small] (rect23) at ($(rect2) + (-0.2*\spacing,-0.35*\spacing)$) {MPI or OpenMP};
        \node[draw, rounded corners, minimum width=0.9*\rectwidth, minimum height=1.1*\rectheight, line width=\linewidth, align=center, font=\small] (rect3) at ($(rect1.east) + ( \spacing, -1.2*\spacing)$) {};
        \node[draw, minimum width=0.5*\rectwidth, minimum height=0.2*\rectheight, line width=\linewidth, align=center, font=\small] (rect31) at ($(rect3) + (-0.2*\spacing, 0.35*\spacing)$) {Par. Algorithms};
        \node[draw, minimum width=0.5*\rectwidth, minimum height=0.2*\rectheight, line width=\linewidth, align=center, font=\small] (rect32) at ($(rect3) + (-0.2*\spacing, 0.1*\spacing)$) {Par. Algorithms};
        \node[draw, minimum width=0.5*\rectwidth, minimum height=0.2*\rectheight, line width=\linewidth, align=center, font=\small] (rect33) at ($(rect3) + (-0.2*\spacing,-0.35*\spacing)$) {Par. Algorithms};
        \node[above, align=center, font=\bfseries\normalsize] at ($(rect1.north) + (0, 0.05*\spacing)$) {User provided model policies}; 
        \node[above, align=center, font=\bfseries\normalsize] at ($(rect2.north) + (0, 0.05*\spacing)$) {Original implementation};
        \node[above, align=center, font=\bfseries\normalsize] at ($(rect3.north) + (0, 0.05*\spacing)$) {Accelerator-friendly implementation};
        \node[below, align=center, font=\itshape\small] at ($(rect2.south) + (0, -0.05*\spacing)$) {Run on multi-core CPU clusters};
        \node[below, align=center, font=\itshape\small] at ($(rect3.south) + (0, -0.05*\spacing)$) {Run on heterogenous systems \\ e.g. GPU-accelerated clusters};
        \draw[-stealth, ultra thick] (rect1.south) -- ($(rect1.south) + (0, -0.5*\spacing)$);
        \draw[-stealth, ultra thick] ($(rect2.east) + (0.2*\spacing, 0.2*\spacing)$) -- node[above, font=\normalsize\itshape] {Automatic instantiation} ($(rect3.west) + (-0.2*\spacing, 0.2*\spacing)$);
        \draw[stealth-stealth, ultra thick] ($(rect2.east) + (0.2*\spacing, -0.2*\spacing)$) -- node[above, font=\normalsize\itshape] {Hybrid execution} ($(rect3.west) + (-0.2*\spacing, -0.2*\spacing)$);
        \draw[-stealth, thick] ($(rect21.east) + (0.05*\spacing,0.0)$) to[out=300,in=60] node[right, font=\normalsize] {MPI} ($(rect22.east) + (0.05*\spacing,0.0)$);
        \draw[-, thick] ($(rect22.east) + (0.05*\spacing,0.0)$) to[out=300,in=90] node[right, font=\normalsize] {\,MPI} ($(rect22.east) + (0.08*\spacing,-0.15*\spacing)$);
        \draw[dotted,-, thick] ($(rect22.east) + (0.08*\spacing,-0.15*\spacing)$) -- node[right, font=\normalsize] {} ($(rect23.east) + (0.08*\spacing,0.15*\spacing)$);
        \draw[-stealth, thick] ($(rect23.east) + (0.08*\spacing,0.15*\spacing)$) to[out=90,in=60] node[right, font=\normalsize] {\,MPI} ($(rect23.east) + (0.05*\spacing,0.0)$);
        \draw[dotted,-, thick] ($(rect22.south) + (0,-0.05*\spacing)$) -- ($(rect23.north) + (0,0.05*\spacing)$);
        \draw[-stealth, thick] ($(rect31.east) + (0.05*\spacing,0.0)$) to[out=300,in=60] node[right, font=\normalsize] {MPI} ($(rect32.east) + (0.05*\spacing,0.0)$);
        \draw[-, thick] ($(rect32.east) + (0.05*\spacing,0.0)$) to[out=300,in=90] node[right, font=\normalsize] {\,MPI} ($(rect32.east) + (0.08*\spacing,-0.15*\spacing)$);
        \draw[dotted,-, thick] ($(rect32.east) + (0.08*\spacing,-0.15*\spacing)$) -- node[right, font=\normalsize] {} ($(rect33.east) + (0.08*\spacing,0.15*\spacing)$);
        \draw[-stealth, thick] ($(rect33.east) + (0.08*\spacing,0.15*\spacing)$) to[out=90,in=60] node[right, font=\normalsize] {\,MPI} ($(rect33.east) + (0.05*\spacing,0.0)$);
        \draw[dotted,-, thick] ($(rect32.south) + (0,-0.05*\spacing)$) -- ($(rect33.north) + (0,0.05*\spacing)$);
    \end{tikzpicture}
    \caption{Illustration of the framework developed to (1) create accelerated data structures (\texttt{AcceleratedLattice}) from Palabos's original object-oriented data structures (\texttt{MultiBlockLattice}), and (2) use both of them in an hybrid fashion.}
    \label{fig:framework}
\end{figure}
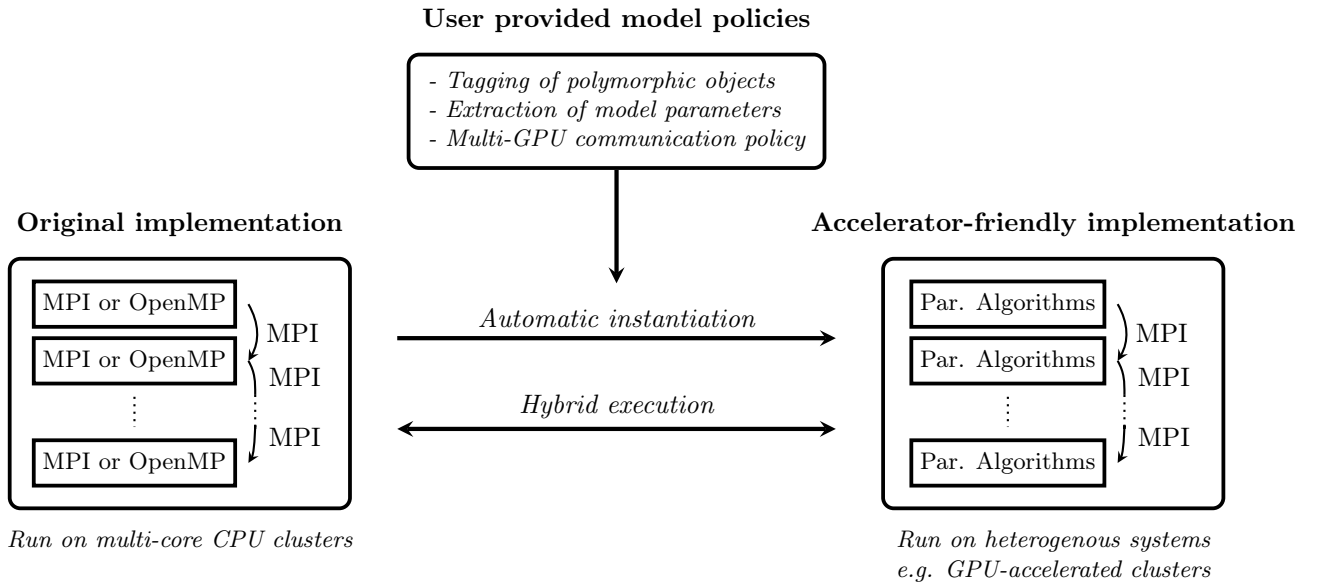
Figure~\ref{fig:framework} illustrates the relationship between these two ways to offer Palabos users per-cell customization of the collision model. It is possible for a GPU-based Palabos application to rely solely on the \texttt{AcceleratedLattice}, which occupies the right space on the figure. Alternatively, the user may set up the simulation on CPU, using the original \texttt{MutliBlockLattice} (left part of the image) and copy all the data, including the specification of the collision models, to the accelerator-friendly container when the initialization is complete. This is especially useful to rapidly port existing Palabos end-user applications to GPU. Then, during program execution, both parallel containers can seemlessly exchange data, including dynamically adapted collision models. Interestingly, the backend which facilitates this exchange represents the major part of the work spent into formulating the Palabos GPU port. While this backend automatizes the work of porting all existing collision models to GPU, an additional manual work is required for every model in the form of user-provided model policies that provide meta-information on the model (central part of Figure~\ref{fig:framework}), for the extraction of model parameters and serialization of the data in view of MPI communication. This work has been performed for many models available in Palabos but is yet to be done by Palabos users who wish to run their own custom models to GPU.

\subsubsection*{GPU kernel size reduction}\label{sec:kernel-size-reduction}
As an undesired side-effect of using an integer-tag approach to handle the polymorphic behavior of nodes in the \texttt{AcceleratedLattice}, the GPU kernel executed for the collision-streaming cycle can reach a considerable size, as it must contain the code for all models that have been assigned a tag. The occupation of GPU memory by this kernel impacts the performance of the code considerably, and unnecessarily, as a typical simulation only uses a limited number of models. This leads to a sizeable performance impact in the current version of the code, as approximately 30 collision models have been ported, and a further performance loss is expected from the ongoing implementation efforts.

To circumvent this problem, the collision-streaming function of the \texttt{AcceleratedLattide} was adapted to allow hand-picking the models that are compiled into the GPU kernel, with the help of template metaprogramming techniques and the possibility of C++ to offer variadic template parameters. The following code extract illustrates the approaches using both a complete kernel (all collision models) and a custom kernel (hand-picked collision models):
 
\begin{verbatim}
    // Plain function call, includes all collision models in the kernel.
    acceleratedLattice.collideAndStream();
    
    // Custom function call with handpicked collision models.
    acceleratedLattice.collideAndStream (
        CollisionKernel<T,DESCRIPTOR,
            CollisionModel::NoDynamics,  // For empty nodes
            CollisionModel::TRT,         // For standard bulk nodes
            CollisionModel::BounceBack,  // For no-slip solid nodes
            CollisionModel::Boundary_RegularizedVelocity_0_1__TRT, // Exterior boundaries
            CollisionModel::Boundary_RegularizedVelocity_0_M1__TRT>() );
\end{verbatim}
In this example, taken from the Berea sandstone benchmark in the validation section, the two final collision models are composite, and are automatically instantiated by Palabos to implement the inlet and outlet boundary conditions. The example shows that it is not always simple for Palabos users to identify the name of all collision models needed by the kernel. Also, missing models cannot be spotted by the compiler and need to be extracted through meticulous work from linker error messages. As a workaround, Palabos offers a function called \texttt{showTemplateArguments(MultiBlockLattice const\&)} which identifies all required model names through introspection of a \texttt{MultiBlockLattice} and produces a Palabos code with the corresponding custom-kernel call to \texttt{collideAndStream} function. In practice, a new Palabos end-user application therefore needs a first run to retrieve the proposer code of the collision-streaming function call. After inclusion of the proposed code, the application can be compiled for the production run.

\subsubsection*{GPU-GPU communication using pinned memory and MPI}
\label{sec:inter-gpu-communication}

\refthree{%
Multi-GPU performance on modern accelerator nodes for memory-bound solvers is dominated by halo exchange efficiency, making direct intra-node GPU-to-GPU communication a prerequisite for parallel scalability. On the NVIDIA DGX~A100 platform used in this work, NVLink/NVSwitch provides a high-bandwidth interconnect (up to 600~GB/s aggregate GPU-to-GPU bandwidth~\cite{nvidia_ucx_multinode}), which is sufficient for the relatively small data volumes exchanged during LBM halo updates, though still lower than on-device GPU memory bandwidth. Any fallback to host memory introduces CPU–GPU transfers with bandwidths orders of magnitude smaller, severely degrading performance. To avoid this, halo buffers must be pinned to the GPU memory, allowing MPI to operate directly on device-resident data without intermediate copies.
}

\refthree{%
Concretely, this is achieved using a \emph{CUDA-aware MPI} implementation, which can send and receive GPU buffers directly. Open~MPI defines CUDA-aware support as the ability for the MPI library to operate on device memory, typically via UCX as the transport layer~\cite{openmpi_cuda_aware}. In UCX-based stacks, intra-node GPU-to-GPU transfers use CUDA inter-process communication mechanisms (\texttt{cuda\_ipc}) to achieve high-performance peer-to-peer transfers across the GPU fabric~\cite{choi2021gpu_ucx,temucin2021nvlink_ucx}.
}



\refthree{
While Palabos primarily relies on ISO~C++ standard parallelism (\texttt{stdpar}) for portability, a small non-standard extension is necessary to achieve optimal GPU-GPU communication. Standard C++ containers such as \texttt{std::vector} allocate pageable host memory, which forces implicit CPU–GPU memory copies during MPI communication, hence severely limiting parallel speedup even at the scale of a single GPU node. To address this, halo buffers are allocated explicitly as CUDA device memory (\texttt{cudaMalloc}), ensuring that MPI receives device pointers and can leverage its CUDA-aware fast paths. Following best practices~\cite{LATT_NVIDIA_BLOG_2_2022,openmpi_cuda_aware}, this workaround is restricted to communication buffers, and it is expected to become unnecessary as C++ and MPI provide improved heterogeneous memory support in the near future.
}

\section{Validation\label{sec:validation}}
This section presents validations of the accuracy and performance of the GPU execution of selected benchmarks cases, all of which are available in the latest release of the Palabos software. 
These benchmark applications differ from classical Palabos end-user applications through the use of the data-oriented \texttt{AcceleratedLattice} in place of the object-oriented \texttt{MultiBlockLattice}, or where appopriate, the simultaneous use of both data containers to warrant a hybrid execution model. The quality of the GPU port is investigated hereafter through several CFD problems of increasing complexity. Being a good compromise between accuracy, robustness and efficiency, all simulations are performed using the D3Q19 lattice -- other velocity discretizations will be included in future releases. Depending on the test requirements, appropriate collision models will be chosen among those already available in Palabos. If not otherwise stated, the convective scaling is adopted so that both the Reynolds and Mach numbers remain unchanged when moving from one mesh resolution to another~\cite{KRUGER_Book_2017}. Solid and inlet boundary conditions are enforced through Dirichlet conditions based on the standard and modified bounce-back conditions respectively~\cite{LADD_JFM_271_1994a,LADD_JFM_271_1994b}. All simulations are performed using single precision, while improving the floating-point accuracy by striping off the constant offset from the populations, as described in the seminal work by Skordos~\cite{SKORDOS_PRE_49_1993}.

\subsection{Transitional flow: Taylor-Green vortex \label{subsec:tgv}}

The transition from a laminar state to a turbulent regime is very common in an airflow, as its kinematic viscosity is of the order of $10^{-5} [m^2/s]$. Hence, it is crucial for aerodynamics simulations to capture such transitions accurately. The Taylor-Green vortex is one of the simplest tests to investigate this phenomenon since it neither requires complex initial state nor boundary conditions~\cite{BRACHET_JFM_130_1983} and is therefore frequently used to benchmark the accuracy of CFD solvers~\cite{DEBONIS_AIAA_382_2013,DAIRAY_JCP_337_2017,JAMMY_JoCS_36_2019,SUSS_CF_257_2023}.

This test consists of the interaction of several vortices that are initialized in a periodic box using the following analytical formulas:
\begin{align}
p &= p_{\infty} + \frac{\rho_{\infty} u_{\infty}}{16}\big[\cos(2z) + 2 \big] \big[\cos(2x) + \cos(2y)\big]\\
u_x &= \phantom{-} u_{\infty} \sin(x) \cos(y) \cos(z)\\
u_y &= - u_{\infty} \cos(x) \sin(y) \cos(z) \\
u_z &= 0
\end{align}
As the simulation starts, the vortices interact and generate smaller vortices through a turbulent energy cascade. This continues until the smallest vortices reach the Kolmogorov scale, at which point they are dissipated through viscous effects.

In this work, $p_{\infty} = \rho_{\infty} c_s^2$, where $\rho_{\infty}=1$ and $c_s^2=1/3$ in lattice units. Furthermore, we will focus on the configuration at $\mathrm{Re}=1600$ for which DNS results are available~\cite{LAIZET_DATASET_TGV_2019}. The Mach number is fixed to $\mathrm{Ma}=0.2$ at all resolutions, which means that the increase of resolution is imposed with convective scaling. To investigate the laminar-to-turbulent transition and the turbulence decay, it is common to monitor the time evolution of the kinetic energy $k$ and the enstrophy $\epsilon$ averaged over the entire domain. In our case, these quantities were computed as follows:
\begin{equation}
k = \frac{1}{\Omega} \int_{\Omega}\frac{u^2}{2}\,\mathrm{d}\Omega,\quad \epsilon = \frac{1}{\Omega} \int_{\Omega}\frac{\omega^2}{2}\,\mathrm{d}\Omega,
\end{equation}
where the size of the simulation domain $\Omega$ is $[2\pi L]^3$ with $L$ the number of points in each direction. The vorticity $\bm{\omega} = \bm{\nabla} \times \bm{u}$ is calculated using an \emph{eighth-order} centered finite difference approximation not to introduce numerical dissipation during post-processing.

This benchmark is performed using a large number of collision models: the single-relaxation-time BGK, multi-relaxation-time collision applied in either raw (RM), Hermite (HM), central (CM), central Hermite (CHM) or cumulant (K) spaces, and the recursive regularized (RR) approach. While the first model relies on the second-order weighted equilibrium, the others are all based on the fourth-order non-weighted equilibrium, detailed in Eqs. (H15)-(H21) of our previous work~\cite{COREIXAS_PRE_100_2019}. To improve the robustness of LBMs in under-resolved conditions, relaxation times related to the bulk viscosity and higher-order moments are fixed to $1.0$ for all collision models except BGK -- see Refs.~\cite{COREIXAS_RSTA_378_2020,LATT_PLOSONE_16_2021} for the exact methodology. 

\begin{figure}[tb]
    \centering
    \includegraphics[width=.99\textwidth]{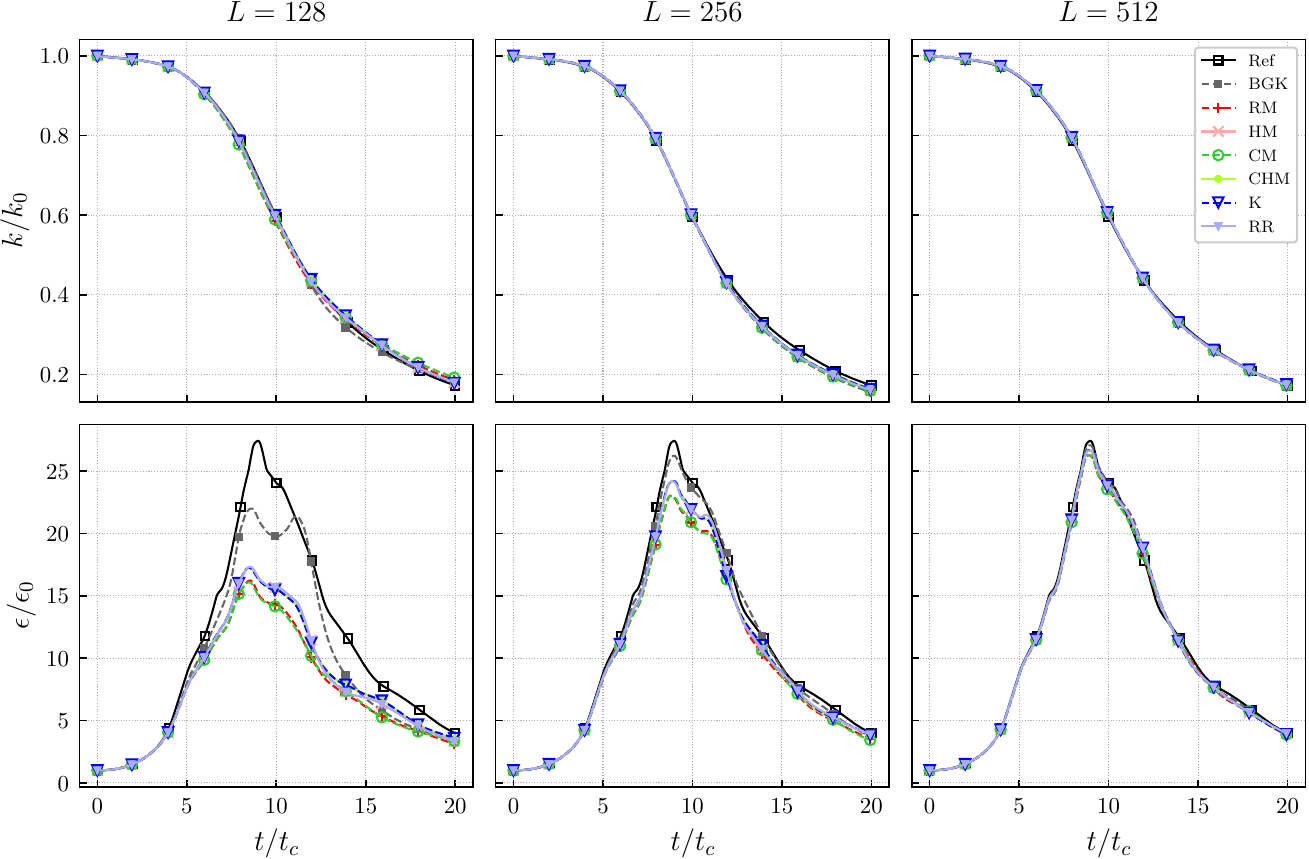}
    \caption{Taylor-Green vortex simulations for $\mathrm{Re}=1600$, $\mathrm{Ma}=0.2$, various mesh resolutions $L\in\{128,256,512\}$, and most collision models implemented in Palabos. Time evolution of the normalized kinetic energy $k/k_0$ (top) and enstrophy $\epsilon/\epsilon_0$. The subscript $0$ stands for quantities at $t=0$, and the convective time is defined as $t_c=2\pi L/ u_{\infty}$.}
    \label{fig:tgv_validation}
\end{figure}

We report in Figure~\ref{fig:tgv_validation} results obtained for increasing mesh resolutions $L\in\{128,256,512\}$ and the above-mentioned collision models. As expected, all models lead to the accurate estimation of the kinetic energy and enstrophy evolution over time for $L=512$. By progressively reducing the mesh resolution from $L=512$ to $128$, the collision step introduces more numerical dissipation. This notably leads to underestimating the enstrophy peak at $t\approx8 t_c$, with the convective time being defined as $t_c=2\pi N/ u_{\infty}$. Since BGK introduces the least amount of numerical dissipation~\cite{MARIE_JCP_228_2009,WISSOCQ_PRE_102_2020,SUSS_CF_257_2023}, it naturally leads to the most accurate results. However, low numerical dissipation implies stability issues for coarse mesh resolutions ($L \leq 128$), hence the unphysical second enstrophy peak observed at $t\approx 12 t_c$, as the BGK model reaches its stability limit for $L \leq 128$. 

Figure~\ref{fig:tgv_validation} presents results obtained for the aforementioned collision models, and increasing mesh resolutions $L\in\{128,256,512\}$. As expected, all models accurately estimate the kinetic energy and enstrophy evolution over time for $L=512$. However, as the mesh resolution is progressively reduced from $L=512$ to $128$, the collision step introduces increasing numerical dissipation. This leads to a notable underestimation of the enstrophy peak at $t\approx8 t_c$, where the convective time is defined as $t_c=2\pi L/ u_{\infty}$. Since BGK introduces the least numerical dissipation~\cite{MARIE_JCP_228_2009,WISSOCQ_PRE_102_2020,SUSS_CF_257_2023}, it naturally produces the most accurate results. However, low numerical dissipation can cause stability issues at coarse mesh resolutions. This results in an unphysical second enstrophy peak at $t\approx 12 t_c$, as the BGK model reaches its stability limit for $L \leq 128$.

\subsection{Wall-bounded flow: Lid-driven cavity \label{subsec:dlc}}

Wall-bounded flows are prevalent in many engineering applications, such as aerodynamics, where the flow over aircraft surfaces needs to be accurately simulated, and in hydraulics, where the behavior of fluid in pipelines and around obstacles is of particular interest.

Hereafter, such flows are used to validate the implementation of no-slip and moving boundary conditions in the GPU version of Palabos. Among them, the lid-driven cavity flow is a perfect candidate as it involves simulating a viscous fluid flow enclosed in a cubic cavity of size $L$~\cite{LATT_PLOSONE_16_2021,PRASAD_PoF_1_1989,ALBENSOEDER_JCP_206_2005,LERICHE_PoF_12_2000,LERICHE_JSC_27_2006,HEGELE_PRE_98_2018,FERRARI_PoF_36_2024}. 
In this setup, all walls are fixed except for the top lid, which moves at a constant velocity $u_x^{(z=L)}=c_s \mathrm{Ma}$. The flow inside the cavity exhibits various characteristics depending on the Reynolds number. At lower Reynolds numbers, the flow tends to be steady and laminar. As the Reynolds number increases, the flow becomes unsteady, displaying a complex interplay of boundary layers, vortices, recirculation zones, and sometimes vortex shedding.

We consider three flow configurations of increasing complexity: steady flow at $\mathrm{Re}=1000$, unsteady laminar flow at $\mathrm{Re}=3200$, and unsteady turbulent flow at $\mathrm{Re}=10000$. For $\mathrm{Re}\leq 3200$, we use a mesh resolution of $L=256$ cells and the standard BGK collision model with second-order weighted equilibrium. For the turbulent case, we adopt the more robust RR collision model and set the relaxation time related to bulk viscosity to unity to ensure stability with the coarse resolution of $L=400$ cells. All simulations are conducted at a Mach number of $\mathrm{Ma}=0.1$.

\begin{figure}[hbt]
\centering
\includegraphics[width=.95\textwidth]{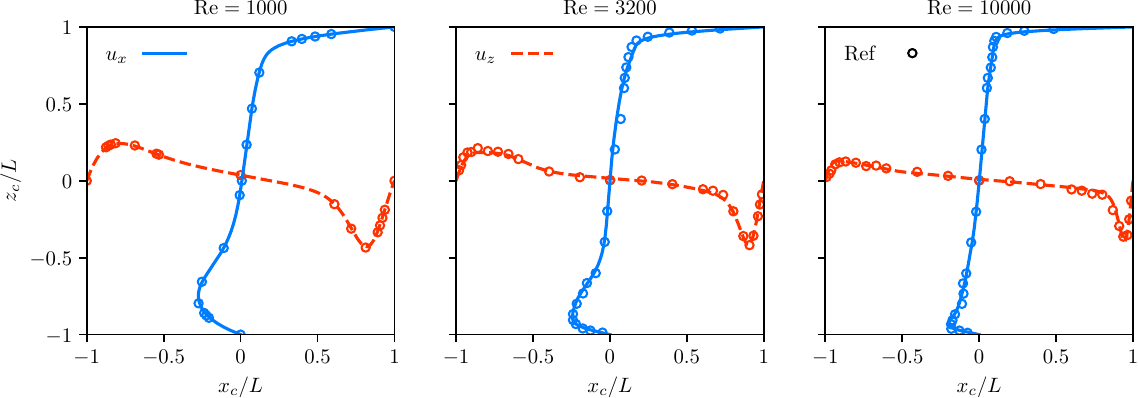}
\caption{Lid-driven cavity flow simulations at various Reynolds numbers and a fixed Mach number of $0.1$.
Results are plotted along the two center lines of the cavity (slice $y=L/2$) using normalized coordinates $\alpha_c/L = 2 \alpha/L - 1$ with $\alpha = x$ or $z$. 
From left to right panels: steady results at $\mathrm{Re}=1000$, non-stationary laminar flow at $\mathrm{Re}=3200$, and turbulent flow at $\mathrm{Re=10000}$.
Reference data (symbols) come from the spectral code~\cite{ALBENSOEDER_JCP_206_2005} and experimental data~\cite{PRASAD_PoF_1_1989} for $Re = 1000$ and $Re\geq 3200$ respectively.}
\label{fig:velocity_profiles_cavity}
\end{figure}

Velocity profiles along the centerlines of the cavity (plane $y=L/2$) are shown in Figure~\ref{fig:velocity_profiles_cavity}. For $\mathrm{Re}=1000$, the flow is steady, and measurements taken after the flow has stabilized (around $78t_c$) match perfectly with simulation data from a spectral code~\cite{ALBENSOEDER_JCP_206_2005}.
The flow becomes unsteady when the Reynolds number increases to $\mathrm{Re}=3200$. Therefore, velocity profiles must be averaged before any comparison with reference data. Averaging is performed every convective time unit ($t_c=L/c_s\mathrm{Ma}$) from $t=50t_c$ to $t=250t_c$. The resulting averaged profiles are in excellent agreement with experimental results~\cite{PRASAD_PoF_1_1989}.
The flow becomes more chaotic at the highest Reynolds number ($\mathrm{Re}=10000$), as the complex interplay between vortices and boundary layers emerges. Hence, velocity profiles are averaged over a longer period, from $t=150t_c$ to $t=500t_c$, with the same output frequency. This final comparison also shows a very good match with the experimental data~\cite{PRASAD_PoF_1_1989}. Consequently, these results confirm the accurate and robust implementation of bounce-back-type boundary conditions in the GPU version of Palabos.

\subsection{Pore-scale fluid flow through porous rocks: Berea sandstone \label{subsec:berea}}

Accurate characterization of fluid flow through porous rocks is essential for determining geological properties such as permeability. Numerical simulations support applications including enhanced oil recovery optimization~\cite{GERRITSEN_ARFM_37_2005,SATTI_SPE_190113_2018,FAGER_E3S_366_2023} and groundwater transport modeling~\cite{XU_CES_52_1997,KIECKHEFEN_ARCBM_11_2020,RETTINGER_JFM_932_2022}. Advances in micro-CT imaging have led to larger and higher-resolution digital samples, demanding significant computational resources for fluid dynamics analysis. 

Building on the extensive use of the original CPU-based version of Palabos in this field (see, e.g., Refs.~\cite{PARMIGIANI_JFM_686_2011,SANEMATSU_JPSE_135_2015,LIU_ATE_93_2016,KELLY_AWR_95_2016,DAVUDOV_IJCG_220_2020,IBRAHIM_ACS_6_2021,NOVITSKA_JPM_28_2025}, among others), the new multi-GPU implementation provides a scalable and efficient solution to address the growing computational demands of high-resolution porous media simulations.
To demonstrate its capabilities, we perform pore-scale simulations of single-phase flow through a Berea sandstone sample --a well-characterized porous medium. The sample, referred to as sample B by Dong et al.\cite{DONG_PRE_80_2009}, was acquired via CT scanning at a resolution of $\delta x = 5.345,\mu\text{m}$ per voxel on a $400 \times 400 \times 400$ grid, yielding a porosity of 19.6\% (the full dataset is available at~\cite{BEREA_V2_2014}).

This case provides a valuable test for the GPU-ported version of Palabos, as the heterogeneous domain presents unique challenges for GPU computation. It also demonstrates hybrid CPU/GPU workflows, where pre- and post-processing remain CPU-based due to their current complexity. Additionally, the benchmark highlights the application of velocity- and pressure-driven boundary conditions using regularized conditions~\cite{LATT_PRE_77_2008}, extending beyond the standard bounce-back approach at the inlet and outlet.

\begin{figure}[hbt]
\centering
\includegraphics[width=.5\textwidth]{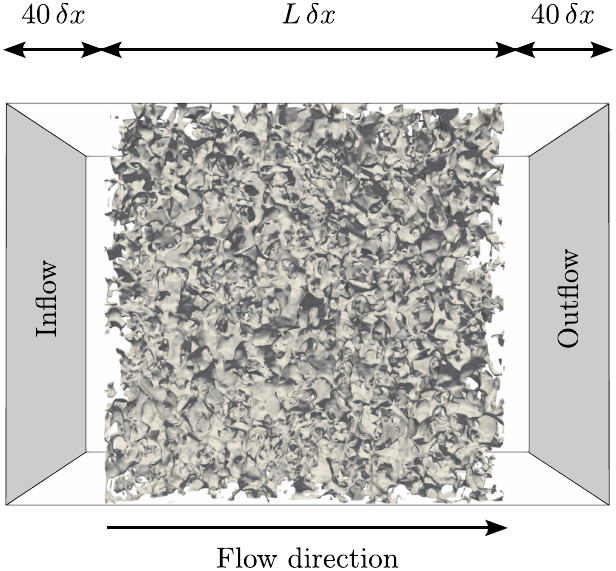}
\caption{Fluid flow through Berea sandstone. Illustration of the simulation domain, with the porous rock at its center, and upstream/downstream subdomains. The simulation is performed with the TRT collision model, and a mesh resolution of $L=400$ using double precision.}
\label{fig:berea}
\end{figure}

As shown in Figure~\ref{fig:berea}, the simulation domain spans $(L+80)\times L \times L$ cells, partitioned into upstream ($40\times L \times L$), porous rock ($L\times L \times L$), and downstream ($40\times L \times L$) sections. Inflow and outflow conditions are applied at the upstream and downstream regions, respectively. The mesh resolution matches the digital sandstone dataset, with $L = 400$ cells. The simulation is validated by computing the absolute permeability:
\begin{equation}
k = \frac{\bar{u}\cdot\nu\cdot lx}{\Delta P},
\end{equation}
where $\bar{u}$ is the mean velocity along the flow direction, $\nu$ the kinematic viscosity, $lx = L,\delta x$ the porous medium length, and $\Delta P$ the pressure drop across the domain. Permeability is reported in millidarcies ($1,\text{mD} = 9.869233 \times 10^{-16},\text{m}^2$). Two approaches are employed: pressure boundaries at inlet and outlet, and velocity boundaries, both using regularized conditions~\cite{LATT_PRE_77_2008}. Simulations proceed to steady state, after which either the pressure drop (velocity boundaries) or average flow rate (pressure boundaries) is measured. Average flow is computed within the porous rock, including zero velocity for solid nodes, though comparable results are obtained using upstream or downstream measurements. The D3Q19-TRT lattice Boltzmann model with double precision is used for fluid nodes, while solid nodes apply a standard bounce-back condition.

\begin{figure}[hbt]
    \centering
    \includegraphics[width=\textwidth]{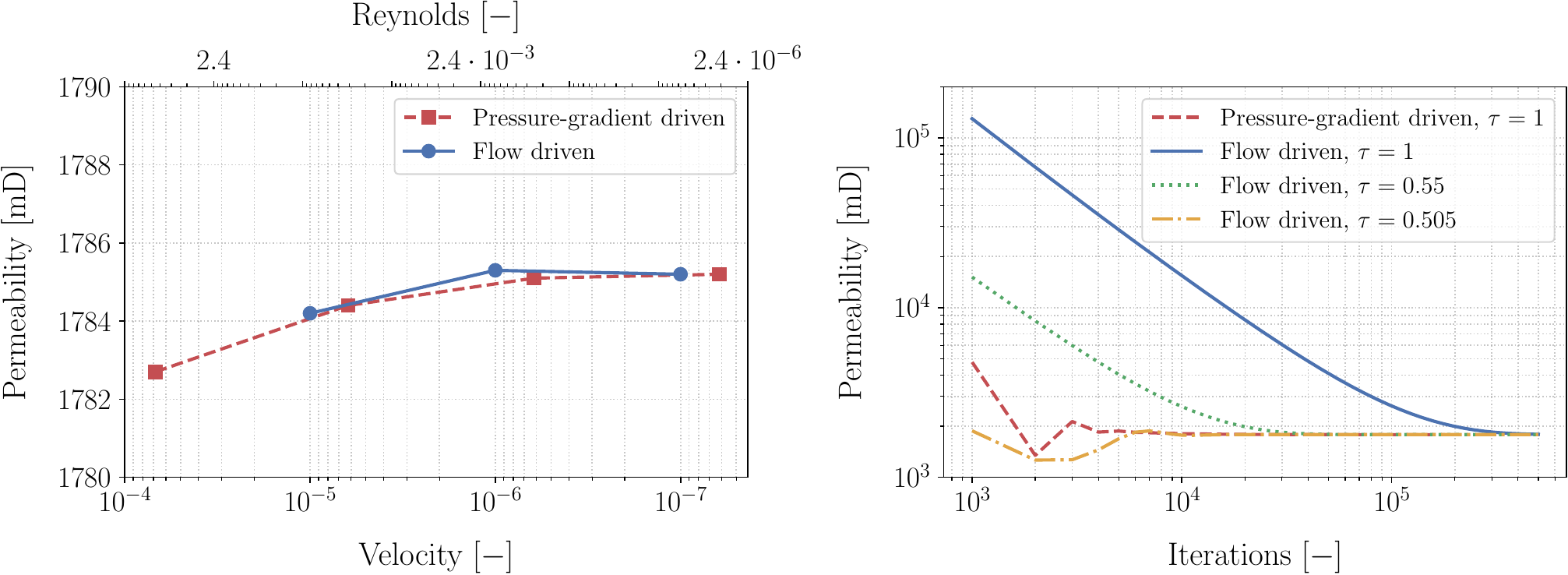}

    \caption{Permeability computed at fixed relaxation time $\tau=1$. Left panel: Result with pressure and velocity condition for permeability vs. velocity and Reynolds number. Right panel: Convergence of permeability at a fixed Reynolds $Re = 0.0024$ with different boundary conditions and numerical parameters.}
    \label{fig:berea_permeability}
\end{figure}

The left panel of Figure~\ref{fig:berea_permeability} presents the permeability results obtained with both boundary condition approaches at decreasing Reynolds numbers. The relaxation time is fixed at $\tau = 1$, while the lattice velocity $u_{lb}$ is progressively reduced to minimize inertial effects and approach the Stokes flow regime. This also reduces the Mach number, mitigating compressibility and finite time step errors. Both methods yield a converged permeability of 1785.2 mD, with agreement to all five reported digits. Even at the highest Reynolds number simulated, the permeability barely deviates from the converged value by less than a percent.

The simulated permeability of 1785.2 mD can be compared with the 1360 mD estimated by pore-network modeling in the work by Dong et al.~\cite{DONG_PRE_80_2009}, representing a difference of 31.2\%.

This agreement should be considered satisfactory, given the well-known challenges in estimating the permeability of Berea sandstone, which is highly sensitive to both sample variability and numerical method details. As demonstrated by Kang et al.~\cite{Kang2019Stokes}, permeability can vary by several orders of magnitude even for samples with similar porosities. This sensitivity is further confirmed in our results: modifying the domain size $L$ causes only slight changes in porosity but leads to significant variations in the measured permeability (see Table~\ref{tab:berea_data}).

\begin{table}[h!]
\centering
\begin{tabular}{c c c c}
\toprule
\multirow{2}{*}{$L$} & \multirow{2}{*}{Porosity (\%)} & \multicolumn{2}{c}{Permeability (mD)} \\
\cmidrule(lr){3-4}
 & & LBM & Pore-network \\
\midrule
300 & 20.41 & 1702 & -- \\
350 & 19.94 & 1820 & -- \\
400 & 19.64 & 1785 & 1360 \\
450 & 19.79 & 1443 & -- \\
500 & 19.86 & 1482 & -- \\
\bottomrule
\end{tabular}
\caption{Berea sandstone properties obtained via a pore-network approach (Dong et al., 2009) and present LBM simulations. The LBM results show that permeability can vary significantly, even for similar porosity values, depending on the domain size $L$.}
\label{tab:berea_data}
\end{table}

Another objective of this type of simulation is to reach steady state as fast as possible to speedup the prediction of permeability. Since LBM is explicit in time, it inherently requires time-dependent evolution to achieve stationarity. However, it is also possible to optimize numerical parameters to accelerate convergence. The right panel of Figure~\ref{fig:berea_permeability} illustrates that through the permeability convergence over time iterations for both pressure-gradient-driven and velocity-driven flows at $\mathrm{Re} = 0.0024$, plotted on logarithmic axes due to the slow convergence behavior. For velocity-driven simulations, the time step (and corresponding Mach number) is systematically reduced by an order of magnitude, while the relaxation time $\tau$ is adjusted to maintain a constant Reynolds number. This tuning is feasible primarily because the TRT collision model decouples wall positioning from $\tau$; without this, time-step convergence of the permeability would not be as straightforward, especially with the BGK collision model.

The results indicate that for $\tau = 1$, the pressure-gradient-driven case reaches five-digit accuracy of the permeability within 20'000 iterations, whereas the velocity-driven case initially requires up to 400'000 iterations. However, further reduction of the lattice velocity and relaxation time enables the velocity-driven case to match the convergence rate of the pressure-gradient approach. In both cases, the permeability of the $400 \times 400 \times 400$ Berea sandstone sample is determined in \emph{under 10 minutes using a single A100-SXM4 (40GB) GPU}.

\section{Performance study\label{sec:perfo}}

\refthree{
Hereafter, we analyze the performance of the multi-GPU version of Palabos and compare it to the same code running on CPU. The GPU and multi-GPU benchmarks were performed on a NVIDIA DGX A100 workstation equipped with four A100-SXM4 (40~GB) GPUs, interconnected via NVLink~3 in a fully connected topology. Each GPU delivers a theoretical memory bandwidth of 1.555~TB/s, while the NVLink 3 interconnect provides up to 600~GB/s of aggregate GPU-to-GPU bandwidth~\cite{nvidia_ucx_multinode}. GPUs were operated at default application clocks with a 400~W power limit and ECC enabled. 
All GPU benchmarks were compiled with NVIDIA HPC Compiler 25.3 using the \texttt{-stdpar} option for GPU offloading, \texttt{-O3} and \texttt{-DNDEBUG} for production optimization, and \texttt{-Msingle -Mfcon} for improved single-precision performance. CPU benchmarks were performed on a dual-socket Intel Xeon Gold 6240R system (48 cores in total) running at 4~GHz, compiled with GCC v13.3 using \texttt{-O3} and \texttt{-DNDEBUG} optimization flags.
}

\subsection{Performance model and memory usage}

Building a performance model is the first step toward understanding how close the implementation of a numerical method is to achieving its peak performance. In the context of GPU acceleration, the available computational power commonly exceeds the one of a CPU by several orders of magnitude. Because of this, the collision step of LBMs is almost instantaneous, and most of the simulation time is spent reading LB data from memory and writing it back after the streaming step. This is why LBMs are commonly referred to as \emph{memory-bound} solvers when running on cluster-class GPUs~\cite{LATT_PLOSONE_16_2021,HOLZER_IJHPCA_2021,HOLZER_PhD_2025,COREIXAS_ARXIV_04465_2025}.

For memory-bound solvers, peak performance can be estimated using the bandwidth of GPUs. For the A100-SXM4 (40GB), the maximum bandwidth is $B_{\mathrm{A100}}$=1.555 GB/s according to the specifications provided by NVIDIA. Dividing $B_{\mathrm{A100}}$ by the amount of data accessed during one time iteration $n_t$, one ends up with the maximum number of cells (or lattices) that can be processed per second. Since this number is very high for LBMs, performance is commonly measured in terms of millions or billions of lattices updated per second, known as MLUPS or GLUPS.

For D3Q19-LBMs, Palabos accesses the 19 populations from a vector $f_i^{in}$ ($i\in\{0,...,18\}$) and writes them back to another vector $f_i^{out}$, in a two-population implementation fashion. Additionally, two flags are used to (1) identify the location of LB data in the Cartesian mesh and (2) check whether the current cell requires boundary condition treatment. Based on this, $n_t = 2 \cdot 19 \cdot n_{\mathrm{store}} + 8 + 4$ Bytes, with $n_{\mathrm{store}}$ being 4 or 8 Bytes depending on whether populations are stored in single or double precision, respectively. This leads to the following peak performance 
\begin{equation}
    P^{\mathrm{peak,sp}}_{\mathrm{A100}} = B_{\mathrm{A100}}/n_t^{\mathrm{sp}} = 9.481\:\mathrm{GLUPS} \quad \mathrm{and} \quad P^{\mathrm{peak,dp}}_{\mathrm{A100}} = B_{\mathrm{A100}}/n_t^{\mathrm{dp}}= 4.921\:\mathrm{GLUPS}
    \label{eq:peak_perfo}
\end{equation}
for single precision (sp) and double precision (dp), respectively.

Memory usage, $mem$, is another important factor that impacts the performance of LBMs when running on GPUs (see Figure 7 in Ref.~\cite{LATT_PLOSONE_16_2021}, and Section 7 of Ref.~\cite{COREIXAS_ARXIV_04465_2025}). Generally speaking, the memory usage can be estimated from the ratio between the memory size and the amount of data stored in memory. For D3Q19-LBMs implemented in Palabos, we obtain for a cubic domain
\begin{equation}
    mem = \dfrac{40\cdot 10^9}{n_t\cdot L^3},
    \label{eq:memory_usage}
\end{equation}
since the memory size of the A100-SXM4 GPU used in this work is 40GB.
Assuming $L=500$, the memory usage is then $mem \approx 51\%$ and $99\%$ for single precision and double precision storage, respectively. In practice, Eq.~(\ref{eq:memory_usage}) can be reverted to compute the minimal mesh resolution to perform simulations at optimal performance, as well as, the maximal domain size that can fit the entire memory of the GPU.

Hereafter, the performance model and memory usage are used to better understand performance metrics of the multi-GPU version of Palabos. As performance can slightly vary from one run to another, each measurement is performed three times. The data provided below then consists of the averaged values obtained from these three measurements.

\subsection{Single GPU performance: Impact of memory usage and collision model\label{subsec:perfo_single_gpu}}

\begin{figure}[htb]
\centering
\includegraphics[width=.95\textwidth]{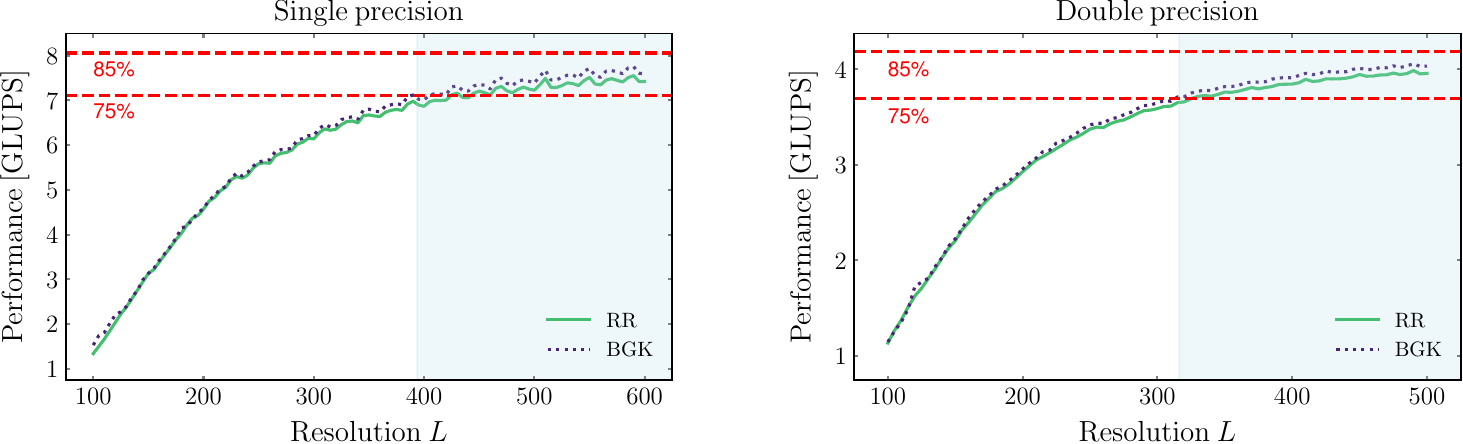}
\caption{Mesh resolution impact on the performance obtained on a single A100-SXM4 (40GB) GPU for the Taylor Green vortex simulation. Results are shown for single (left panel) and double (right panel) precision computations, as well as for BGK and RR collision models. The light blue area indicates a memory usage of at least 25\% of the total memory capacity of the A100-SXM4 (40GB). The red dashed lines represent 75\% and 85\% of peak performance.}
\label{fig:perfo_mesh_impact}
\end{figure}

Let us start with a performance study based on the Taylor-Green vortex benchmark presented in Section~\ref{subsec:tgv}. Figure~\ref{fig:perfo_mesh_impact} illustrates the impact of mesh resolution, and the corresponding memory usage, on performance using BGK and RR collision models. As observed in a previous studies~\cite{LATT_PLOSONE_16_2021,COREIXAS_ARXIV_04465_2025}, performance is an increasing function of the mesh resolution, and the corresponding memory load of the A100 GPU. It exceeds 75\% of peak performance when memory usage is greater than 25\% (light blue area), and it gradually approaches 80-85\% of peak performance as memory usage increases. This trend is consistent for both the simple BGK and the more robust RR collision models, and the peak performance value is close to what one gets with GPU-native approaches~\cite{HOLZER_IJHPCA_2021,HOLZER_PhD_2025}.

\begin{figure}[htb]
\centering
\includegraphics[width=.95\textwidth]{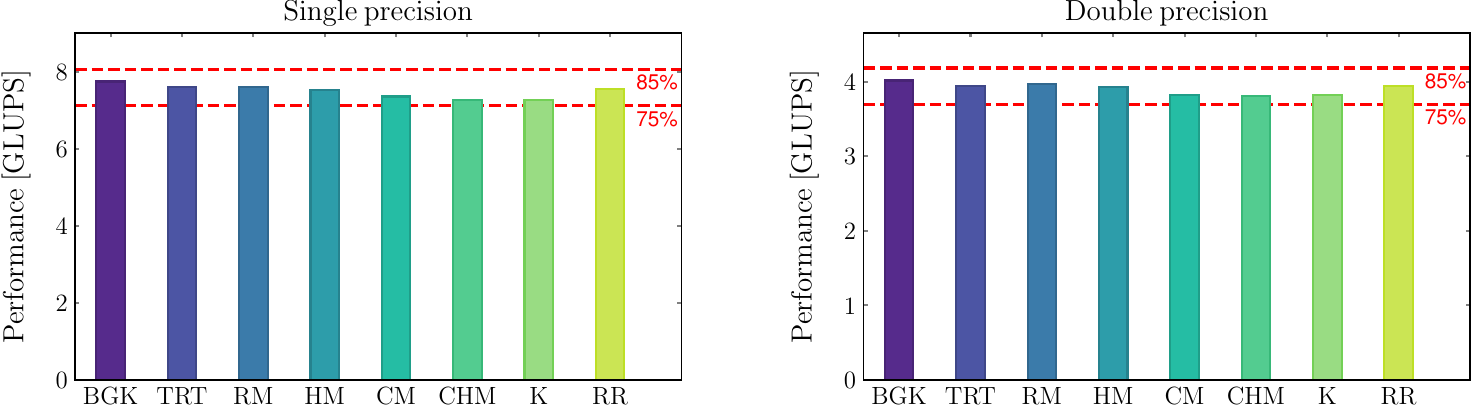}
\caption{Collision model impact on the performance obtained on a single A100-SXM4 (40GB) GPU for the Taylor Green vortex simulation. Simulations are performed using $L=590$ and $490$ when single and double precision are used, respectively. The red dashed lines stands for 75 and 85\% of the peak performance.}
\label{fig:perfo_comp_collision}
\end{figure}

In order to check the impact of collision models used in Section~\ref{sec:validation} on the performance of Palabos, we compare them in Figure~\ref{fig:perfo_comp_collision}. The latter shows that all D3Q19-LBMs achieve between 75\% and 85\% of peak performance. This indicates that the collision step has minimal impact on performance, provided that (1) the GPU offers sufficient computational power, and (2) collision models are optimized to minimize computations. Both aspects are crucial for gaming-class GPUs, which have significantly reduced computational power for double precision arithmetic. As such, a number of optimization were implemented in Palabos, and they involve unrolling loops, using symmetry rules to reduce the number of terms to compute, and precomputing redundant terms. The interested reader may refer to Section 2.6 in Ref~\cite{LATT_PLOSONE_16_2021} and Section 4.3 in Ref~\cite{HENNIG_SIAM_45_2023} for more details on optimization strategies of the collision step.

\subsection{Multi-GPU performance: Weak and strong scaling analysis\label{subsec:perfo_scalings}}

\begin{figure}[tb]
\centering
\includegraphics[width=.95\textwidth]{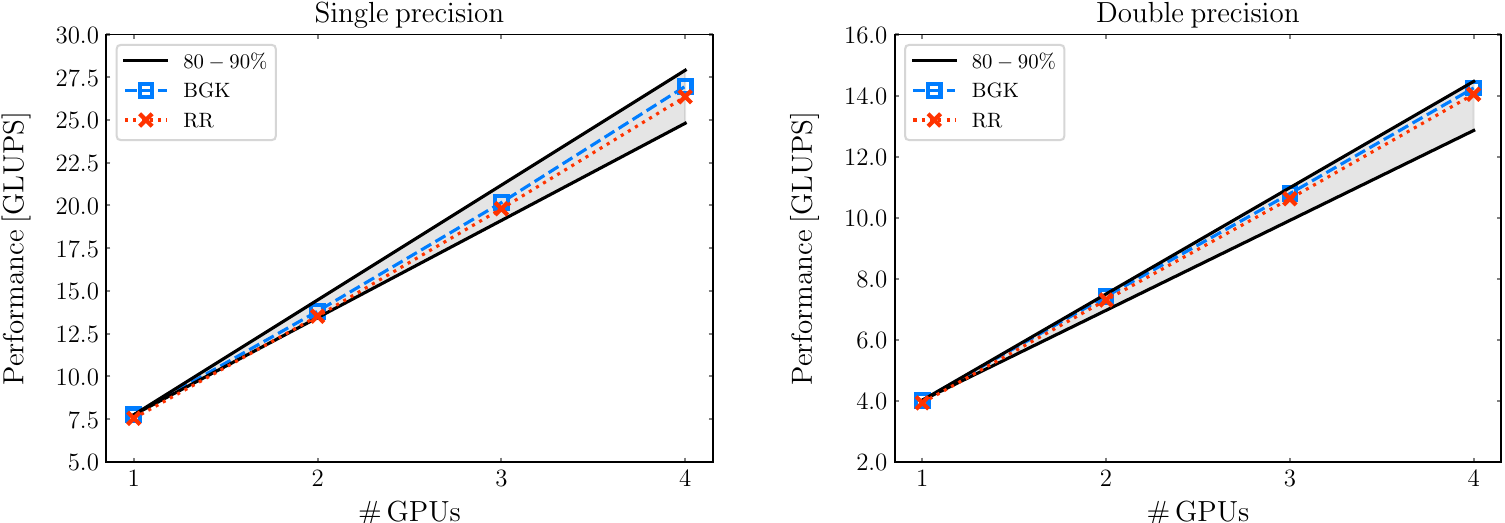}
\caption{Weak scaling benchmark performed on the Taylor Green vortex case using a full DGX-A100-SXM4 (4$\times$40GB). The simulation domain length is $L\in\{590,743,851,937\}$ and $L\in\{480,604,692,761\}$ for single and double precision, respectively. Performance remains between 80 and 90\% of the ideal weak scaling (grey area).}
\label{fig:perfo_weak_scaling}
\end{figure}

\begin{figure}[tb]
\centering
\includegraphics[width=.95\textwidth]{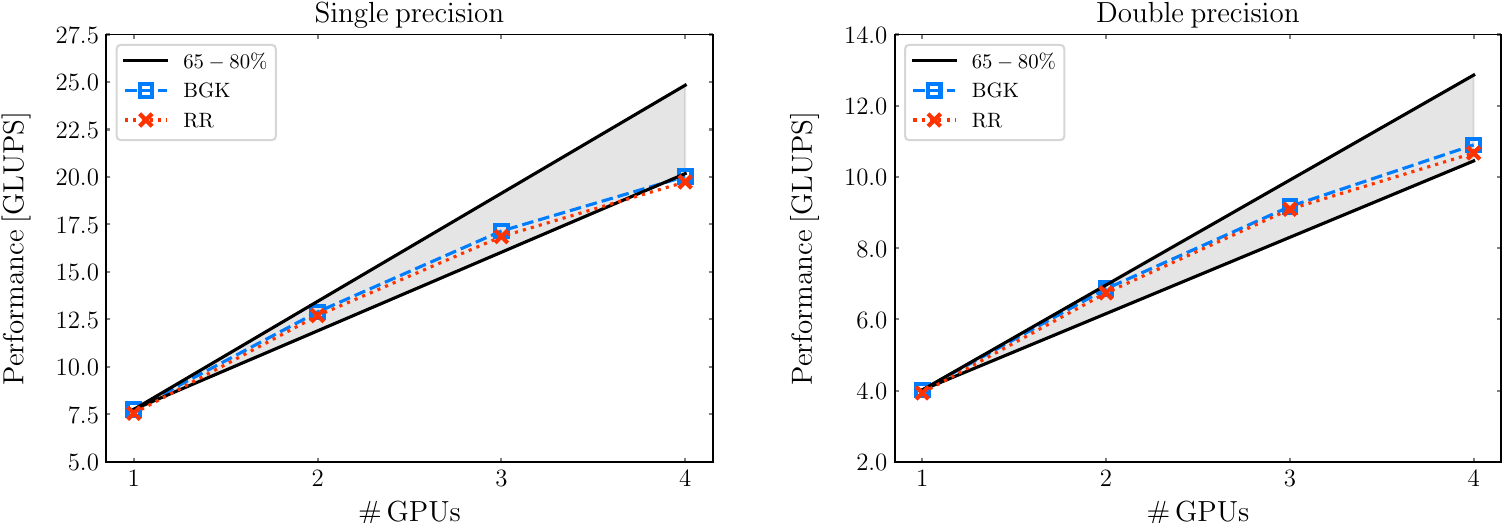}
\caption{Strong scaling benchmark performed on the Taylor Green vortex case using a full DGX-A100-SXM4 (4$\times$40GB). The simulation domain length is fixed to $L = 590$ and $L = 480$ for single and double precision, respectively. Performance remains between 65 and 80\% of the ideal strong scaling (grey area).}
\label{fig:perfo_strong_scaling}
\end{figure}

Below, we aim to quantify the performance of our multi-GPU porting through two specific configurations: weak scaling and strong scaling. The weak scaling study evaluates how effectively the numerical method handles larger problems while maintaining a constant amount of data processed per GPU. Conversely, the strong scaling analysis measures how the simulation time is reduced when the problem size remains fixed, but the number of GPUs increases.

As with the single GPU performance analysis, we use the Taylor-Green vortex case to first evaluate the weak scaling of Palabos on a full DGX-A100-SXM4 (4$\times$40GB). To meet the requirements of a weak scaling study, the simulation domain length $L$ is chosen to (1) end up with optimal performance on a single GPU, and (2) ensure each GPU processes approximately the same number of cells. Choosing $L=590$ for single precision and $L=490$ for double precision satisfies the first point, while leaving a safety margin to avoid fully loading the GPU. The second point is achieved by multiplying $L$ by $\sqrt[3]{2}$, $\sqrt[3]{3}$ and $\sqrt[3]{4}$ for two, three and four GPUs, respectively. Corresponding results are gathered in Figure~\ref{fig:perfo_weak_scaling}. 
It is interesting to note that Palabos experiences minimal performance loss when increasing the number of GPUs, even without overlapping communications and computations. Furthermore, the measured weak scaling is linear which means the performance is not deteriorated when the full DGX is used. This result holds true for the BGK, RR and other collision models (not shown here), hence, corroborating the findings presented in Section~\ref{subsec:perfo_single_gpu}. One possible reason for the favorable linear weak scaling of Palabos is that the ratio of communicated to processed data remains relatively constant across all configurations.

In the strong scaling analysis, we start with an optimal domain size and increase the number of GPUs from one to four. Consequently, the effective domain size processed per GPU is $L_{\mathrm{eff}}=\sqrt[3]{L^3/\#\mathrm{GPUs}} < L$. Practically speaking, $L_{\mathrm{eff}}\in\{590,468,409,372\}$ and $L_{\mathrm{eff}}\in\{480,381,333,302\}$ for single and double precision when $\#\mathrm{GPUs}\in\{1,2,3,4\}$. Referring to Figure~\ref{fig:perfo_mesh_impact}, one can expect a degradation in performance per GPU due to a reduced memory usage as the number of GPUs increases. This non-linear effect is further expected to add up with the performance loss (10-20\%) induced by the non-overlapped data communication, and already observed in the weak scaling study. Figure~\ref{fig:perfo_strong_scaling} presents strong scaling data for the BGK and RR collision models. The results are consistent for both D3Q19-LBMs and confirm the predicted non-linear deterioration in performance. Quantitatively speaking, both loss effects result in a strong scaling efficiency of about 80\% for two GPUs and 65\% for the full DGX. These remain particularly impressive results for the first multi-GPU release of Palabos, especially considering the complexity and size of the codebase, which exceeds 500,000 lines of code, and the fact that similar trends have been observed when comparing weak and strong scaling properties of GPU-native LBM solvers~\cite{HOLZER_PhD_2025}.

\subsection{Performance comparison across simulation cases\label{subsec:perfo_scalings_comp_simu}}

\begin{figure}[h!]
\centering
\includegraphics[width=.99\textwidth]{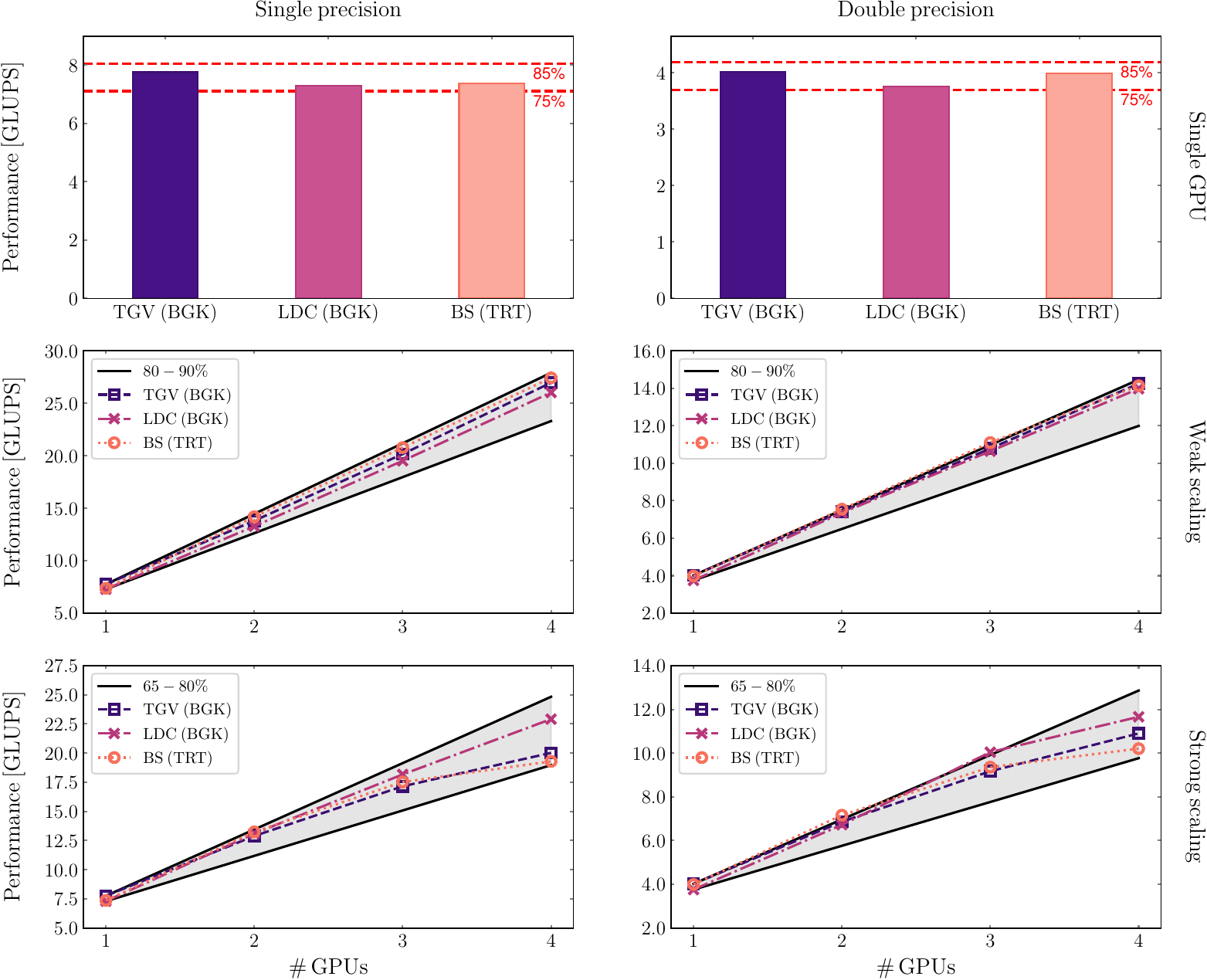}
\caption{Performance comparison for all configurations considered in Section~\ref{sec:validation}: Taylor-Green vortex (TGV), lid-driven cavity (LDC), and Berea sandstone (BS). Corresponding simulations are performed using BGK, BGK and TRT collision models. Single GPU, weak scaling, and strong scaling results are gathered in the top, middle and bottom row, respectively. Single GPU performance is measured using $L \in\{590, 590, 400\}$ for single precision, and $\{480, 480, 317\}$ for double precision. For weak and strong scaling, the simulation domain sizes are computed according to the rules described in Section~\ref{subsec:perfo_scalings}.}
\label{fig:perfo_comp_multi_simu}
\end{figure}

In this final study, we evaluate the ability of the multi-GPU version of Palabos to handle various configurations in an efficient manner. Notably, we are interested in quantifying the impact of boundary conditions on performance: periodic for the Taylor-Green vortex (TGV), standard/modified bounce-back for the lid-driven cavity (LDC), and regularized/bounce-back/periodic for the Berea sandstone (BS). Additionally, the BS case will help us evaluate the robustness of the multi-GPU implementation of Palabos in handling heterogeneous treatment of contiguous data in memory, especially inside the sandstone where switching between solid and fluid dynamics is very frequent.

We start with data obtained running the TGV (BGK, $L=590$), the LDC (BGK, $L=590$), and the BS (TRT, $L=400$) cases on a single GPU for single precision. Corresponding results are presented in the top row of Figure~\ref{fig:perfo_comp_multi_simu}.
It can be seen that performance is similar for all cases as it remains between 75 and 85\% of the peak performance, which is a solid performance, compatible with the one obtained in simpler LBM codes based on C++ parallel algorithms~\cite{LATT_PLOSONE_16_2021}. \refthree{Moreover, the achieved performance is comparable to that of other open-source, state-of-the-art GPU LBM codes. For uniform-mesh applications on A100-SXM4 (40~GB) GPUs, and using the peak performance~\ref{eq:peak_perfo} as a reference value for this GPU, WaLBerla (CUDA backend) reports a performance of approximately 85\% of peak~\cite{HOLZER_PhD_2025}, while FluidX3D (OpenCL backend) achieves around 84\%~\cite{lehmann2022accuracy}. Both values are near the upper bound of the range plotted on Figure~\ref{fig:perfo_comp_multi_simu}, which confirms the efficiency of the Palabos GPU port based on modern C++.}


In the weak and strong scaling analyses, discrepancies are observed between the TGV, LDC, and BS cases, as illustrated in the middle and bottom rows of Figure~\ref{fig:perfo_comp_multi_simu}.
For all configurations, the weak scaling is almost linear, remaining between 80\% and 90\% of its ideal value. As previously noted for the TGV case (Section~\ref{subsec:perfo_scalings}), the strong scaling is non-linear and ranges from 65 to 80\% of its ideal value.  

Going into more details, performance discrepancies between configurations are reduced when using double precision, especially in the weak scaling analysis. As expected, periodicity introduces additional communications, leading to more significant performance degradation for the TGV case, compared to LDC and BS, as the number of GPUs increases. Despite the heterogeneous treatments of fluid and solid cells, both LDC and BS perform well, with a slight advantage for LDC. However, the strong scaling performance of BS deteriorates significantly when utilizing three or more GPUs. This decline is primarily due to the lower memory usage resulting from the reduced mesh resolution, which was necessary to stay within the memory limits of the GPU card, as additional memory is required to store the geometry of the sandstone.

In conclusion, these results demonstrate that by utilizing C++ parallel algorithms to port Palabos to GPU, it is possible to achieve near-peak performance on a single A100-SXM4 (40GB) GPU and very good weak and strong scaling on a DGX-A100-SXM4 (4$\times$40GB). This is especially encouraging given that computations and communications have not been overlapped in this first release of the multi-GPU version of Palabos fluid solver.

\section{Conclusion\label{sec:conclusion}}
In this work, we have presented the GPU porting of the widely used lattice Boltzmann software library Palabos, using only the standard parallelism available in modern C++. This approach avoids any reliance on external libraries or vendor-specific extensions, which guarantees portability and makes the code easier to maintain across different hardware platforms. The new hybrid software architecture connects the customizable, object-oriented data structures of the CPU version with data-oriented structures optimized for the GPU. This allows existing features to be gradually migrated to the GPU without duplicating large amounts of code.

It is important to emphasize that Palabos has never been limited to being just a simulation tool. From its beginning, it has been designed as a framework for scientists and engineers to develop, test, and improve their own lattice Boltzmann models and algorithms. 
This is why this foundational philosophy has been meticulously followed throughout the GPU port of its fluid solver.
Consequently, the design abstracts the complexity of GPU programming from the end user, offering clear and well-structured mechanisms for implementing new and potentially complex LBM models. Users can thus focus on model development without requiring in-depth knowledge of GPU architectures.

The GPU version of Palabos fluid solver has been validated on several demanding test cases, including the laminar-to-turbulent transition of a Taylor-Green vortex, flow in a lid-driven cavity, and pore-scale flow through a Berea sandstone sample. In all cases, the results show very good to excellent agreement with reference data. Performance tests show that the GPU port achieves near-peak performance on a single GPU and good scaling on multi-GPU systems,
even without overlapping communication and computation.
This demonstrates that the current porting strategy provides performance comparable to GPU-native codes, while maintaining the flexibility and accessibility that have made Palabos widely used in the scientific community.

The experience gained in this work also serves as a practical example of how complex C++ scientific software can be adapted to GPU execution while keeping the code both understandable and accessible to a wide range of users. Current efforts are focused on porting additional LBM velocity sets (e.g., D3Q7, D3Q15, D3Q27, D3Q39), as well as \texttt{MultiScalarField} objects that are commonly used for the coupling of LBM and finite difference schemes. Future developments include adding overlap of communication and computation to improve multi-GPU performance, extending support to other types of accelerator hardware, and further improving the ease with which users can create new simulations and models. With these advancements, Palabos is ready to remain a leading open-source framework for advanced lattice Boltzmann simulations on current and future high-performance computing systems.

\section*{Acknowledgements}
The authors would like to express their gratitude to the NVIDIA team for their support and assistance during the initial stages of the GPU porting of the Palabos fluid solver using modern C++. Special thanks are extended to Gonzalo Brito and Jeff Larking for fruitful conversations, technical support, and contributions to the NVIDIA developer two-part article.
C.C. acknowledges financial support from BNBU Research Fund $\mathrm{n^o UICR0700094}$-$24$ entitled ``GPU-accelerated multidisciplinary methods for industrial computational fluid dynamics''.

\section*{Declaration of generative AI and AI-assisted technologies in the writing process}
During the preparation of this work the authors revised numerous sentences and paragraphs using the online user interfaces of DeepL Write and ChatGPT 4 to enhance grammar, style, and readability. ChatGPT 4 was also used as a productivity tool to produce the \LaTeX{} source of some of the sketches. After using these services, the authors reviewed and edited the content as needed and take full responsibility for the content of the publication.

\end{document}